\documentclass[aps,showpacs,nofootinbib,superscriptaddress]{revtex4}

\usepackage{graphicx}
\usepackage{dcolumn}

\def\slashchar#1{\setbox0=\hbox{$#1$}
   \dimen0=\wd0 \setbox1=\hbox{/} \dimen1=\wd1
   \ifdim\dimen0>\dimen1 \rlap{\hbox to \dimen0{\hfil/\hfil}} #1
   \else  \rlap{\hbox to \dimen1{\hfil$#1$\hfil}} / \fi}
\def\tstrut{\vrule height2.5ex depth0pt width0pt} 
\begin{document}
\title{Inclusive Quasi-Elastic Charged--Current Neutrino--Nucleus Reactions.}
\author{J. Nieves} \affiliation{Departamento de F\'\i sica Moderna,
\\Universidad de Granada, E-18071 Granada, Spain} \author{J.E. Amaro}
\affiliation{Departamento de F\'\i sica Moderna, \\Universidad de
Granada, E-18071 Granada, Spain} \author{M. Valverde}
\affiliation{Departamento de F\'\i sica Moderna, \\Universidad de
Granada, E-18071 Granada, Spain}
\begin{abstract}
\rule{0ex}{3ex} The Quasi-Elastic (QE) contribution of the nuclear
inclusive electron scattering model developed in Ref.~\cite{GNO97} is
extended to the study of electroweak Charged Current (CC) induced
nuclear reactions, at intermediate energies of interest for future
neutrino oscillation experiments. The model accounts for,
among other nuclear effects, long range nuclear (RPA) correlations,
Final State Interaction (FSI) and  Coulomb corrections.  Predictions for the
inclusive muon capture in $^{12}$C and the reaction $^{12}$C
$(\nu_\mu,\mu^-)X$ near threshold are also given. RPA correlations are
shown to play a crucial role and their inclusion leads to one of the
best existing simultaneous description of both processes, with
accuracies of the order of 10-15\% per cent for the muon capture rate
and even better for the LSND measurement. 
\end{abstract}


\maketitle



\section{Introduction}

The neutrino induced reactions in nuclei at intermediate energies play
an important role in the study of neutrino properties and their
interaction with matter~\cite{Nuint04}. A good example of this is the
search for neutrino oscillations, and hence physics beyond the
standard model~\cite{Fu98}. Several experiments are planned or under
construction~\cite{Nuint04}, aimed at determining the neutrino
oscillation parameters with high precision. The data analysis will be
sensitive to sources of systematic errors, among them nuclear effects
at intermediate energies (nuclear excitation energies ranging from
about 100 MeV to 500 or 600 MeV), being then of special interest to
come up with an unified Many-Body Framework (MBF) in which the
electroweak interactions with nuclei could be systematically
studied. Such a framework would necessarily include three different
contributions: i) QE processes, ii) pion production and two body
processes from the QE region to that beyond the $\Delta(1232)$
resonance peak, and iii) double pion production and higher nucleon
resonance degrees of freedom induced processes. Any model aiming at
describing the interaction of neutrinos with nuclei should be firstly
tested against the existing data of interaction of real and virtual
photons with nuclei. There exists an abundant literature on this
subject, but the only model which has been successfully compared with
data at intermediate energies and that systematically includes the
first and the second of the contributions, and partially also the
third one, mentioned above, is that developed in Refs.~\cite{CO92}
(real photons), and~\cite{GNO97} (virtual photons). This model is able
to describe inclusive electron--nucleus scattering, total nuclear
photo-absorption data, and also measurements of photo-- and
electro--nuclear production of pions, nucleons, pairs of nucleons,
pion-nucleon pairs, etc. The building blocks of this model are: 1) a
gauge invariant model for the interaction of real and virtual photons
with nucleons, mesons and nucleon resonances with parameters
determined from the vacuum data, and 2) a microscopic treatment of
nuclear effects, including long and short range nuclear
correlations~\cite{OTW82}, FSI, explicit meson and $\Delta (1232)$
degrees of freedom, two and even three nucleon absorption channels,
etc.  The nuclear effects are computed starting from a Local Fermi Gas
(LFG) picture of the nucleus, and their main features, expansion
parameter and all sort of constants are completely fixed from previous
hadron-nucleus studies (pionic atoms, elastic and inelastic
pion-nucleus reactions, $\Lambda-$ hypernuclei, etc.)~\cite{pion}. The
photon coupling constants are determined in the vacuum, and the model
has no free parameters. The results presented in Refs.~\cite{CO92} and
\cite{GNO97} are predictions deduced from the framework developed in
Refs.~\cite{OTW82}--\cite{pion}. One might think that LFG description
of the nucleus is poor, and that a proper finite nuclei treatment is
necessary. For inclusive processes and nuclear excitation energies of
at least 100 MeV or higher, the findings of Refs.~\cite{GNO97},
\cite{CO92} and \cite{pion} clearly contradict this conclusion. The
reason is that in these circumstances one should sum up over several
nuclear configurations, both in the discrete and in the continuum, and
this inclusive sum is almost no sensitive to the details of the
nuclear wave function, in sharp contrast to what happens in the case
of exclusive processes where the final nucleus is left in a determined
nuclear level. On the other hand, the LFG description of the nucleus
allows for an accurate treatment of the dynamics of the elementary
processes (interaction of photons with nucleons, nucleon resonances,
and mesons, interaction between nucleons or between mesons and
nucleons, etc.)  which occur inside the nuclear medium. Within a
finite nuclei scenarious, such a treatment becomes hard to implement,
and often the dynamics is simplified in order to deal with more
elaborated nuclear wave functions. This simplification of the dynamics
cannot lead to a good description of nuclear inclusive electroweak
processes at the intermediate energies of interest for future neutrino
oscillation experiments.

Our aim is to extend the nuclear inclusive electron scattering model
of Ref.~\cite{GNO97}, including the axial CC degrees of freedom, to
describe neutrino and antineutrino induced nuclear reactions. This is
a long range project; in this work we present our model for the QE
region, and hence it constitutes the first step towards this end.  We 
 also present results for the inclusive muon capture in $^{12}$C
and predictions for the LSND measurement of the reaction $^{12}$C
$(\nu_\mu,\mu^-)X$ near threshold. Both processes are clearly
dominated by the QE contribution and are drastically affected by the
inclusion of nuclear correlations of the RPA type. We find
\begin{equation}
\Gamma \left[^{12}_{\mu^-}{\rm C}\right ] = 3.2 \times 10^4 \,{\rm
  s}^{-1} \quad
{\bar \sigma}(\nu_\mu) = 11.9 \times 10^{-40} \,{\rm cm}^{2},
\end{equation}
in a good agreement with data (discrepancies of the order of 10-15\%
for the muon capture rate), despite that those measurements involve
extremely low nuclear excitation energies (smaller than 15-20 [25-30]
MeV in the first [second] case), where the LFG picture of the
nucleus might break down. However, it turns out that the present model
provides one of the best existing combined description of these two
low energy measurements, what increases our confidence
on the QE predictions of the model at the higher transferred
energies of interest for future neutrino experiments. Some preliminary
results were presented in Ref.~\cite{nuint04}.

There exists  an abundant literature both on the inclusive muon
capture in nuclei,~\cite{Pr59}--\cite{Ko03}, and on
the CC neutrino--nucleus cross section in the QE
region~\cite{Ki85}--\cite{Gr04}. Among  all these works, we would
like to highlight those included in Refs.~\cite{Ch90} ($\mu-$capture)
and~\cite{Si92} (CC QE scattering) by Oset and collaborators. The
framework presented in these works  is quite
similar to that employed here. Nicely and in a very simple manner,
these works show the most important features of the strong nuclear
renormalization effects affecting the nuclear weak responses in the QE
region. The main differences with the work presented here concern to
the RPA re-summation, which, and as consequence of the acquired
experience in the inclusive electron scattering studies~\cite{GNO97},
is here improved, by considering effects not only in the vector-isovector
channel of the nucleon-nucleon interaction, but also in the
scalar-isovector one. Besides, a more complete tensorial treatment of
the RPA response function is also carried out in this work, leading
all of these improvements to a better agreement to data.

In addition here, we also evaluate the FSI effects for intermediate
nuclear excitation energies, not taken into account in the works of
Ref.~\cite{Si92}, on the neutrino induced nuclear cross sections.

The paper is organized as follows. After this introduction, we deduce
the existing relation among the CC neutrino inclusive nuclear cross
sections and the gauge boson $W$ selfenergy inside  the nuclear
medium (Sect.~\ref{sec:neu}). In Sects.~\ref{sec:self}
and~\ref{sec:anti} we study in detail the QE contribution to the
neutrino and antineutrino nuclear cross section, paying a special
attention to the role played by the strong renormalization of the CC
in the medium (Sect.~\ref{sec:rpa}) and to the FSI effects
(Sect.~\ref{sec:fsi}). The inclusive muon capture in nuclei and the
relation of this process with  inclusive neutrino induced
 reactions are examined in Sect.\ref{sec:imc}. Results and 
main conclusions of this work are compiled in Sects.~\ref{sec:res}
and~\ref{sec:concl}. Finally, in the  Appendix, 
some detailed formulae are given.

\section{CC  Neutrino Inclusive Nuclear Reactions} \label{sec:neu}
\subsection{General Formulae}
We will focus on the inclusive nuclear reaction driven by the electroweak CC
\begin{equation}
\nu_l (k) +\, A_Z \to l^- (k^\prime) + X 
\label{eq:reac}
\end{equation}
though the generalization of the obtained expressions to antineutrino
induced reactions, neutral current processes, or  inclusive muon
capture in nuclei is straightforward.

The double differential cross section, with respect to the outgoing
lepton kinematical variables,  for the process of Eq.~(\ref{eq:reac})
is given in the Laboratory (LAB) frame by
\begin{equation}
\frac{d^2\sigma_{\nu l}}{d\Omega(\hat{k^\prime})dE^\prime_l} =
\frac{|\vec{k}^\prime|}{|\vec{k}~|}\frac{G^2}{4\pi^2} 
L_{\mu\sigma}W^{\mu\sigma} \label{eq:sec}
\end{equation}
with $\vec{k}$ and $\vec{k}^\prime~$ the LAB lepton momenta, $E^{\prime}_l =
(\vec{k}^{\prime\, 2} + m_l^2 )^{1/2}$ and $m_l$ the energy, and the
mass of the outgoing lepton ($m_\mu = 105.65$ MeV, $m_e = 0.511$
MeV  ), $G=1.1664\times 10^{-11}$ MeV$^{-2}$, the
Fermi constant and $L$ and $W$ the leptonic and hadronic tensors,
respectively. The leptonic tensor is given by (in our convention, we
take $\epsilon_{0123}= +1$ and the metric $g^{\mu\nu}=(+,-,-,-)$):
\begin{eqnarray}
L_{\mu\sigma}&=& L^s_{\mu\sigma}+ {\rm i} L^a_{\mu\sigma} =
 k^\prime_\mu k_\sigma +k^\prime_\sigma k_\mu
- g_{\mu\sigma} k\cdot k^\prime + {\rm i}
\epsilon_{\mu\sigma\alpha\beta}k^{\prime\alpha}k^\beta \label{eq:lep}
\end{eqnarray}
The hadronic tensor includes all sort of non-leptonic
vertices and corresponds to the charged electroweak transitions of the
target nucleus, $i$, to all possible final states. 
It is thus given by\footnote{Note that: (i)
Eq.~(\ref{eq:wmunu}) holds with states normalized so that $\langle
\vec{p} | \vec{p}^{\,\prime} \rangle = (2\pi)^3 2p_0
\delta^3(\vec{p}-\vec{p}^{\,\prime})$, (ii) the sum over final states
$f$ includes an integration $ \int \frac{d^3p_j}{(2\pi)^3 2E_j}$, for
each particle $j$ making up the system $f$, as well as a sum over all
spins involved.}
\begin{eqnarray}
W^{\mu\sigma} &=& \frac{1}{2M_i}\overline{\sum_f } (2\pi)^3
\delta^4(P^\prime_f-P-q) \langle f | j^\mu_{\rm cc}(0) | i \rangle
 \langle f | j^\sigma_{\rm cc}(0) | i \rangle^*
\label{eq:wmunu}
\end{eqnarray}
with $P^\mu$ the four-momentum of the initial nucleus, $M_i=P^2$ 
the target nucleus mass, $P_f^\prime$  the total four momentum of
the hadronic state $f$ and $q=k-k^\prime$ the four momentum
transferred to the nucleus.  The bar over the sum denotes the
average over initial spins, and finally for the CC we take
\begin{equation}
j^\mu_{\rm cc} = \overline{\Psi}_u\gamma^\mu(1-\gamma_5)(\cos\theta_C \Psi_d +
\sin\theta_C \Psi_s) 
\end{equation}
with $\Psi_u$, $\Psi_d$ and $\Psi_s$ quark fields, and $\theta_C$ the
Cabibbo angle ($\cos\theta_C= 0.974$). By construction, the hadronic
tensor accomplishes
\begin{eqnarray}
W^{\mu\sigma}= W^{\mu\sigma}_s + {\rm i} W^{\mu\sigma}_a 
\end{eqnarray}
with $W^{\mu\sigma}_s$ ($W^{\mu\sigma}_a$) real symmetric
(antisymmetric) tensors. To obtain Eq.~(\ref{eq:sec}) we have
neglected the four-momentum carried out by the intermediate $W-$boson
with respect to its mass, and have used  the existing relation between
the gauge weak coupling
constant, $g = e/\sin \theta_W$, and  the Fermi constant:  
$G/\sqrt 2 = g^2/8M^2_W$, with $e$ the electron charge, $\theta_W$ the
Weinberg angle and $M_W$ the $W-$boson mass.

The hadronic tensor is
completely determined by six independent, Lorentz scalar and real,
structure functions $W_i(q^2)$,
\begin{equation}
\frac{W^{\mu\nu}}{2M_i} = - g^{\mu\nu}W_1 + \frac{P^\mu
  P^\nu}{M_i^2} W_2 + {\rm i}
  \frac{\epsilon^{\mu\nu\gamma\delta}P_\gamma q_\delta}{2M_i^2}W_3 +  
\frac{q^\mu  q^\nu}{M_i^2} W_4 + \frac{P^\mu q^\nu + P^\nu q^\mu}
{2M_i^2} W_5+ {\rm i}\frac{P^\mu q^\nu - P^\nu q^\mu}
{2M_i^2} W_6
\end{equation}
Taking
$\vec{q}$ in the $z$ direction, ie, $\vec{q}= |q| {\vec u}_z$, and
$P^\mu = (M_i, \vec{0})$, it is straightforward to  find the six structure
functions in terms of the $W^{00}, W^{xx}=W^{yy}, W^{zz}, W^{xy}$ and $W^{0z}$
components of the hadronic tensor\footnote{ These relations read
\[
W_1=\frac{W^{xx}}{2M_i}, \quad W_2=\frac{1}{2M_i} \left (W^{00}+W^{xx}
+ \frac{(q^0)^2}{|\vec{q}\,|^2}(W^{zz}-W^{xx})-
2\frac{q^0}{|\vec{q}\,|}{\rm Re}~W^{0z}\right), \quad 
W_3=-{\rm i}\frac{W^{xy}}{|\vec{q}\,|}, 
\]
\begin{equation}
 W_4 =\frac{M_i}{2|\vec{q}\,|^2}(W^{zz}-W^{xx}),\quad W_5 = 
\frac{1}{|\vec{q}\,|} \left ( {\rm Re}~W^{0z}-
\frac{q^0}{|\vec{q}\,|}(W^{zz}-W^{xx})\right), \quad 
W_6=\frac{{\rm Im}~W^{0z}}{|\vec{q}\,|} \label{eq:ws}
\end{equation}
}. After contracting with the leptonic tensor we obtain
\begin{eqnarray}
\frac{d^2\sigma_{\nu l}}{d\Omega(\hat{k^\prime})dE^\prime_l} &=&
\frac{|\vec{k}^\prime| E^\prime_l M_iG^2}{\pi^2} \left \{ 2W_1
\sin^2\frac{\theta^\prime}{2} + W_2
\cos^2\frac{\theta^\prime}{2} - W_3 \frac{E_\nu+E^\prime_l}{M_i}
\sin^2\frac{\theta^\prime}{2} +
\frac{m_l^2}{E^\prime_l(E^\prime_l+|\vec{k}^\prime|)}\left
     [W_1\cos\theta^\prime -\frac{W_2}{2}\cos\theta^\prime \right. \right.\nonumber\\
       &+&\left.\left.\frac{W_3}{2}\left(\frac{E^\prime_l
+|\vec{k}^\prime|}{M_i} -
       \frac{E_\nu+E^\prime_l}{M_i} \cos\theta^\prime\right)
+ \frac{W_4}{2}\left(\frac{m^2_l}{M_i^2}\cos\theta^\prime +
       \frac{2E^\prime_l(E^\prime_l+|\vec{k}^\prime|)}{M_i^2}
 \sin^2\theta^\prime\right)-W_5\frac{E^\prime_l+|\vec{k}^\prime|}{2M_i}\right]
\right\} \label{eq:cross}
\end{eqnarray}
with $E_\nu$ the incoming neutrino energy and $\theta^\prime$ the
outgoing lepton scattering angle.   The cross section does not
depend on $M_i$, as can be seen from the relations of
Eq.~(\ref{eq:ws}), and also note that the structure function $W_6$
does not contribute.

\subsection{Hadronic Tensor and the Gauge Boson Selfenergy in the Nuclear
  Medium}

\begin{figure}[b]
\centerline{\includegraphics[height=6cm]{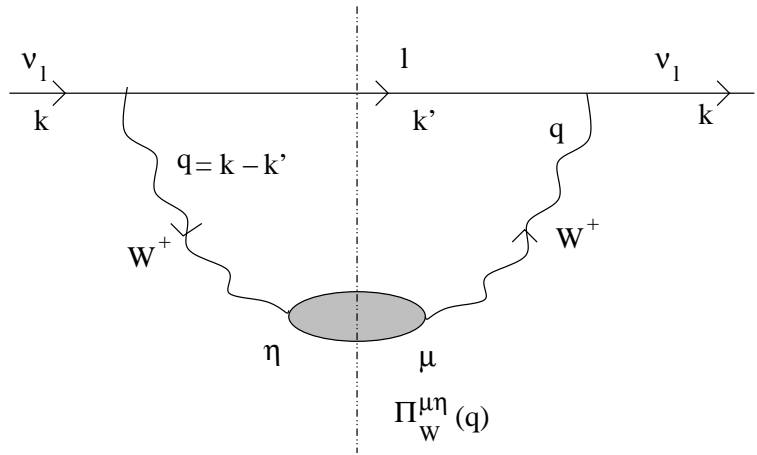}}
\caption{\footnotesize Diagrammatic representation of the neutrino
  selfenergy in nuclear matter. }\label{fig:fig1}
\end{figure}

\begin{figure}[b]
\centerline{\includegraphics[height=12.0cm]{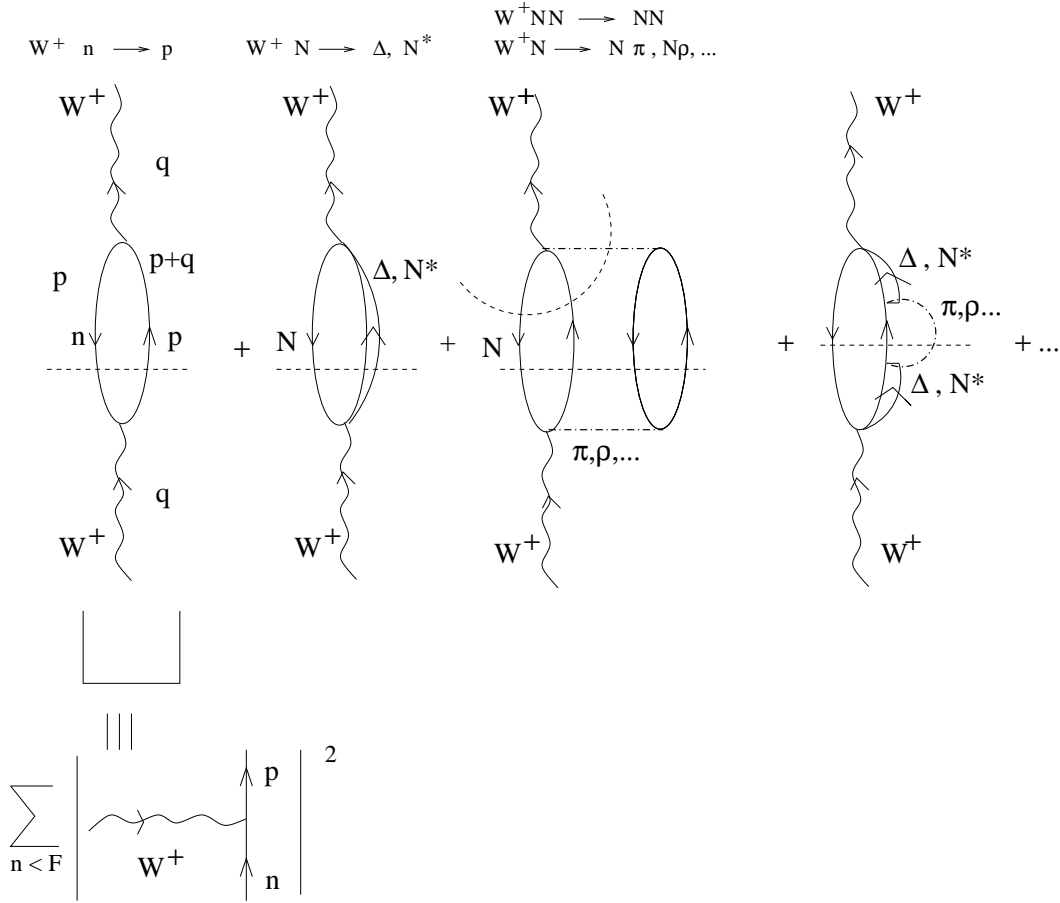}}
\caption{\footnotesize Diagrammatic representation of some diagrams
  contributing to the $W^+-$selfenergy. }\label{fig:fig2}
\end{figure}

In our MBF, the hadronic tensor is determined by
the $W^+-$boson selfenergy, $\Pi^{\mu\rho}_W(q)$, in the nuclear
medium. We follow here the formalism of Ref.~\cite{GNO97}, and we
evaluate the selfenergy, $\Sigma_\nu^r(k;\rho)$, of a neutrino, with
four-momentum $k$ and helicity $r$, moving in infinite nuclear matter
of density $\rho$. Diagrammatically this is depicted in
Fig.~\ref{fig:fig1}, and we get
\begin{equation}
-{\rm i\,}\Sigma_\nu^r(k;\rho) = \int \frac{d^4q}{(2\pi)^4}
 \overline{u}_r(k) \Big \{ -{\rm i} \frac{g}{2\sqrt 2} \gamma^\mu_L\,
 {\rm i}D_{\mu\alpha}(q) \left (-{\rm i}\Pi^{\alpha\beta}_W(q;\rho)
 \right ) \,{\rm i} \, D_{\beta\sigma}(q)
 {\rm i} \frac{\slashchar{k}^\prime + m_l } {k^{\prime 2}-m^2_l+ {\rm
 i}\epsilon} \left (-{\rm i} \frac{g}{2\sqrt 2} \right )
 \gamma^\sigma_L \Big \} u_r(k)
\end{equation}
with $D_{\mu\alpha}(q) = \Big ( -g_{\mu\alpha}+ q_\mu q_\alpha/M^2_W
 \Big )/\Big(q^2-M^2_W + {\rm i }\epsilon\Big) $,
 $\Pi^{\mu\eta}_W(q;\rho)$ is the virtual $W^+$ selfenergy in the medium, 
 $\gamma^\mu_L = \gamma^\mu(1-\gamma_5)$, and  spinor
 normalization given by ${\bar u} u = 2m$. Since right-handed
 neutrinos are sterile, only the left-handed neutrino selfenergy,
 $\Sigma_\nu(k;\rho)$,  is not zero and obviously $\Sigma_\nu(k;\rho)
 = \sum_r \Sigma_\nu^r(k;\rho)$. The sum over helicities leads to
 traces in the Dirac's space a thus we get
\begin{equation}
\Sigma_\nu(k;\rho) = \frac{8{\rm i} G}{\sqrt 2 M^2_W} \int
\frac{d^4q}{(2\pi)^4} \frac{L_{\eta\mu} \Pi^{\mu\eta}_W(q;\rho) 
}{k^{\prime 2}-m^2_l+ {\rm i}\epsilon}
\end{equation}
The neutrino disappears from the elastic flux, by inducing one
particle-one hole (1p1h), 2p2h $\cdots$ excitations, $\Delta(1232)-$hole
($\Delta$h) excitations or creating pions, etc... at a rate given by
\begin{equation}
\Gamma (k;\rho) = - \frac{1}{k^0 }{\rm Im} \Sigma_\nu (k;\rho)
\end{equation}
We get the imaginary part of $\Sigma_\nu$ by using the Cutkosky's
rules. In this case we cut with a straight vertical line (see
Fig.~\ref{fig:fig1}) the intermediate lepton state and those implied
by the $W-$boson polarization (shaded region). Those states are then
placed on shell by taking the imaginary part of the propagator,
selfenergy, etc. Thus, we obtain for $k^0 > 0$
\begin{equation}
{\rm Im} \Sigma_\nu(k) = \frac{8G}{\sqrt 2 M^2_W}\int \frac{d^3
  k^\prime}{(2\pi)^3 }\frac{\Theta(q^0) }{2E^{\prime}_l} 
~ {\rm Im}\left\{ \Pi^{\mu\eta}_W(q;\rho) L_{\eta\mu} \right\} 
\label{eq:ims}
\end{equation}
with $\Theta(...)$ the Heaviside function. Since
$\Gamma dt dS$ provides a probability times a differential of area,
which is a contribution to  the $(\nu_l,l)$ cross section, we have
\begin{eqnarray}
d\sigma &=& \Gamma(k;\rho)dtdS = - \frac{1}{k^0 }{\rm Im} \Sigma_\nu
(k;\rho) dtdS = - \frac{1}{|\vec{k}|} {\rm Im} \Sigma_\nu (k;\rho)
d^3r
\end{eqnarray}
and hence the nuclear cross section is given by
\begin{equation}
\sigma = - \frac{1}{|\vec{k}|} \int {\rm Im} \Sigma_\nu (k;\rho(r)) d^3r
\end{equation}
where we have substituted $\Sigma_\nu$ as a function of the nuclear
density at each point of the nucleus and integrate over the whole
nuclear volume. Hence, we assume the Local Density Approximation, (LDA)
which, as shown in Ref.~\cite{CO92}, is an excellent approximation for
volume processes like the one studied here. Coming back to Eq.~(\ref{eq:ims})
we find 
\begin{eqnarray}
\frac{d^2\sigma_{\nu l}}{d\Omega(\hat{k^\prime})dk^{\prime 0}} &=&-
\frac{|\vec{k}^\prime|}{|\vec{k}~|}\frac{G^2}{4\pi^2}
\left(\frac{2\sqrt 2}{g}\right)^2
 \int\frac{d^{\,3}r}{2\pi} \Big \{   L_{\mu\eta}^s ~{\rm
  Im}(\Pi_W^{\mu\eta}+\Pi_W^{\eta\mu}) 
 -     L_{\mu\eta}^a ~{\rm
  Re}(\Pi_W^{\mu\eta}-\Pi_W^{\eta\mu)} \Big \} \Theta(q^0)
\end{eqnarray}
and then by comparing to Eq.~(\ref{eq:sec}), the hadronic tensor reads
\begin{eqnarray}
W^{\mu\sigma}_s &=& - \Theta(q^0) \left (\frac{2\sqrt 2}{g} \right )^2 
\int \frac{d^3 r}{2\pi}~ {\rm Im}\left [ \Pi_W^{\mu\sigma} 
+ \Pi_W^{\sigma\mu} \right ] (q;\rho)\label{eq:wmunus}\\
W^{\mu\sigma}_a &=& - \Theta(q^0) \left (\frac{2\sqrt 2}{g} \right )^2
 \int \frac{d^3 r}{2\pi}~{\rm Re}\left [ \Pi_W^{\mu\sigma} 
- \Pi_W^{\sigma\mu}\right] (q;\rho) \label{eq:wmunua}
\end{eqnarray}
 As we see, the basic object is the selfenergy of the Gauge Boson
($W^{\pm}$) inside of the nuclear medium. Following the lines of
Ref.~\cite{GNO97}, we should perform a many body expansion, where the
relevant gauge boson absorption modes would be systematically
incorporated: absorption by one nucleon, or a pair of nucleons or even
three nucleon mechanisms, real and virtual meson ($\pi$, $\rho$,
$\cdots$) production, excitation of $\Delta$ of higher resonance
degrees of freedom, etc. In addition, nuclear effects such as RPA or
Short Range Correlations\footnote{For that purpose we use an effective
interaction of the Landau-Migdal type.} (SRC) should also be taken
into account.  Some of the $W-$absorption modes are depicted in
Fig.~\ref{fig:fig2}.
Up to this point the formalism is rather general and its applicability
has not been restricted to the QE region. In this work we will
focus on the QE contribution to the total cross section, and it will
be analyzed in detail in the next section.

\section{QE contribution to $\Pi_W^{\mu\nu}(q;\rho)$}
\label{sec:self}
The virtual $W^+$ can be absorbed by one nucleon leading to the QE
contribution of the nuclear response function. Such a contribution
corresponds to a 1p1h nuclear excitation (first of the diagrams
depicted in Fig.~\ref{fig:fig2}). To evaluate this selfenergy, 
the free nucleon propagator in the medium is required.
\begin{equation}
S(p\,; \rho) = (\slashchar{p}+M) G(p\,; \rho),
\qquad G(p\,; \rho) = \left ( \frac{1}{p^2-M^2+ {\rm
    i}\epsilon}  +  \frac{2\pi{\rm i}}{2E(\vec{p}\,)}
    \delta(p^0-E(\vec{p}\,)) \Theta(k_F-|\vec{p}~|) \right ) \label{eq:Gp}
\end{equation}
with the local Fermi momentum $k_F(r)= (3\pi^2\rho(r)/2)^{1/3}$,
$M=940 $ MeV the nucleon mass, and $E(\vec{p}\,)= \sqrt{M^2 +
\vec{p}^{\,2}}$. We will work on a non-symmetric nuclear matter with
different Fermi sea levels for protons, $k_F^p$, than for neutrons,
$k_F^n$ (equation above, but replacing $\rho /2$ by $\rho_p $ or $\rho_n$,
with $\rho=\rho_p + \rho_n$).  On the other hand,  for the $W^+pn$ vertex we take
\begin{equation}
< p; \vec{p}^{~\prime}=\vec{p}+\vec{q}~ | j^\alpha_{cc}(0) | n;
\vec{p}~> =
\bar{u}(\vec{p}{~^\prime})(V^\alpha-A^\alpha)u(p) 
\end{equation}
with vector and axial  nucleon currents  given by
\begin{equation}
V^\alpha = 2\cos\theta_C \times  \left ( F_1^V(q^2)\gamma^\alpha + {\rm
i}\mu_V \frac{F_2^V(q^2)}{2M}\sigma^{\alpha\nu}q_\nu\right), \qquad
A^\alpha = \cos\theta_C G_A(q^2) \times \left (  \gamma^\alpha\gamma_5 + 
\frac{2M}{m_\pi^2-q^2}q^\alpha\gamma_5 \right) 
\end{equation}
with $m_\pi=139.57$ MeV. Partially conserved axial current and
invariance under G-parity have been assumed to relate the
pseudoscalar form factor to the axial one and to discard a term of the
form $(p^\mu+p^{\prime \mu})\gamma_5$ in the axial sector,
respectively.  Invariance under time reversal guarantees that all form
factors are real. Besides, Due to isospin symmetry, the vector
form factors are related to the electromagnetic ones\footnote{We use
the parameterization of Galster and collaborators~\protect\cite{Ga71}
\begin{equation}
F_1^N = \frac{G_E^N+\tau G_M^N}{1+\tau}, \qquad \mu_N F_2^N =
\frac{G_M^N- G_E^N}{1+\tau}, \quad G_E^p = \frac{G_M^p}{\mu_p}=
\frac{G_M^n}{\mu_n} = -(1+\lambda_n\tau) \frac{G_E^n}{\mu_n \tau} =
\left(\frac{1}{1-q^2/M^2_D}\right)^2 \label{eq:f1n}
\end{equation}
with $\tau=-q^2/4M^2$, $M_D=0.843$ MeV, $\mu_p=2.792847$, $\mu_n=-1.913043$ and
$\lambda_n=5.6$.} 
\begin{equation}
 F_1^V(q^2) =  \frac12 \left (F_1^p(q^2)-F_1^n(q^2)\right),\qquad
 \mu_V F_2^V(q^2) = \frac12 \left ( \mu_p F_2^p(q^2) - \mu_n F_2^n(q^2)\right) 
\end{equation}
and for the axial form-factor we use
\begin{equation} 
G_A(q^2) = \frac{g_A}{(1-q^2/M_A^2)^2},\quad g_A=1.257, \quad M_A =
1.049~ {\rm GeV} \label{eq:axial}
\end{equation}
With all of these ingredients is straightforward to evaluate the 
contribution to the $W^+-$selfenergy  of the
first  diagram of Fig.~\ref{fig:fig2}, 
\begin{eqnarray}
-{\rm i} \Pi_W^{\mu\nu} (q^0,\vec{q}\,) = - \cos^2\theta_C
\left(\frac{g}{2\sqrt 2}\right)^2 \int
 \frac{d^4p}{(2\pi)^4}A^{\mu\nu}(p,q) G(p; \rho_n)G(p+q; \rho_p)
 \label{eq:defPi}
\end{eqnarray}
with the CC nucleon tensor given by
\begin{eqnarray}
A^{\mu\nu}(p,q) &=& Tr\left\{\left(2F_1^V\gamma^\mu - 2{\rm
i}\mu_V \frac{F_2^V}{2M}\sigma^{\mu\alpha}q_\alpha - G_A \left
( \gamma^\mu\gamma_5 - \frac{2M}{m_\pi^2-q^2}q^\mu\gamma_5
\right)\right)(\slashchar{p}+\slashchar{q}+M)\right.\nonumber\\
&\times& \left.\left(2F_1^V\gamma^\nu + 2{\rm
i}\mu_V \frac{F_2^V}{2M}\sigma^{\nu\beta}q_\beta - G_A \left
( \gamma^\nu\gamma_5 + \frac{2M}{m_\pi^2-q^2}q^\nu\gamma_5
\right)\right)(\slashchar{p}+M)\right\} \label{eq:traamunu}
\end{eqnarray}
The Dirac's space traces above can be easily done and the nucleon tensor
can be found in Sect.~\ref{sec:amunu} of the Appendix. Subtracting
the divergent vacuum contribution in Eq.~(\ref{eq:defPi}), we finally
get from Eqs.~(\ref{eq:wmunus}) and~ (\ref{eq:wmunua})
\begin{eqnarray}
W^{\mu\nu}(q^0,\vec{q}\,) 
&=& - \frac{\cos^2\theta_C}{2M^2} \int_0^\infty dr r^2 \Big \{
2 \Theta(q^0) \int \frac{d^3p}{(2\pi)^3}\frac{M}{E(\vec{p})}
\frac{M}{E(\vec{p}+\vec{q})}  
\Theta(k_F^n(r)-|\vec{p}~|) \Theta(|\vec{p}+\vec{q}~|-k_F^p(r))
 \nonumber\\
&\times&  (-\pi)\delta(q^0 +
E(\vec{p}) -E(\vec{p}+\vec{q}~))   A^{\nu\mu}(p,q)|_{p^0=E(\vec{p})~} 
\Big \} \label{eq:res}
\end{eqnarray}
The $d^3p$ integrations above can be analytically done  and all of them
are determined by the imaginary part of the relativistic isospin
asymmetric Lindhard function,
$\overline {U}_R(q,k_F^n,k_F^p)$.  Explicit expressions can be found
in  Sect.~\ref{sec:rel} of the Appendix. 

Up to this point the treatment is fully relativistic and the four
momentum transferred to the nucleus can be comparable or higher than
the nucleon mass\footnote{The only limitation on its  size is
given by possible quark effects, not included in the nucleon
form-factors of Eqs.~(\ref{eq:f1n})--(\ref{eq:axial}).  }. At low and
intermediate energies, RPA effects become extremely large, as shown
for instance in Ref~\cite{Ch90}. To account for RPA effects, we will
use a nucleon--nucleon effective force~\cite{Sp77} determined from
calculations of nuclear electric and magnetic moments, transition
probabilities and giant electric and magnetic multipole resonances
using a non-relativistic nuclear dynamics scheme. This force,
supplemented by nucleon--$\Delta(1232)$ and
$\Delta(1232)$--$\Delta(1232)$ interactions~\cite{OTW82}-\cite{pion},
was successfully used in the work of Ref.~\cite{GNO97} on inclusive
nuclear electron scattering.  In this latter reference a
non-relativistic LFG is also employed. Thus, it is of interest to
discuss also the hadronic tensor of Eq.~(\ref{eq:res}) in the context
of a non-relativistic Fermi gas. This is easily done by replacing the
factors $M/E(\vec{p})$ and $M/E(\vec{p}+\vec{q})$ in
Eq.~(\ref{eq:res}) by one. Explicit expressions can be now found in
Sect.~\ref{sec:nonrel} of the Appendix.

Pauli blocking, through the imaginary part of the Lindhard function, is
the main nuclear effect included in the hadronic tensor of
Eq.~(\ref{eq:res}). In the next subsections, we will study different
nuclear corrections to $W^{\mu\nu}$.

To finish  this section, we devote a few words to the Low Density
Theorem (LDT). At low nuclear densities the
imaginary part of the relativistic isospin
asymmetric Lindhard function can be approximated by
\begin{equation}
{\rm Im}\overline{U}_R(q,k_F^n,k_F^p) \approx - \pi \rho_n
\frac{M}{E(\vec{q}\,)} \delta(q^0+M-E(\vec{q}\,))
\end{equation}
and thus one readily finds\footnote{The energy of the outgoing lepton
  is completely fixed once the Fermi distribution of the nucleons is
  neglected. Thus all structure functions $W_i$ get the energy
  conservation Dirac's delta into their definition. Indeed, we have
\begin{equation}
W^{\mu\nu} =
\frac{N\cos^2\theta_C}{8ME(\vec{q}\,)}\delta(q^0+M-E(\vec{q}\,))
\times A^{\nu\mu}\Big|_{ p=(M,\vec{0})}
\end{equation} }
\begin{equation}
\sigma_{\nu_l +\,
  A_Z \to l^-  + X } \approx N \sigma_{\nu_l  +\, n \to l^-
   + p}, \qquad N=A-Z  \label{eq:ldt}
\end{equation}
which accomplishes with the LDT. For future purposes we
give in Sect.~\ref{sec:free} of the Appendix the $\nu_l +\, n \to l^- + p $
differential cross section.

\subsection{RPA Nuclear Correlations} \label{sec:rpa}
\begin{figure}[b]
\centerline{\includegraphics[height=14.0cm]{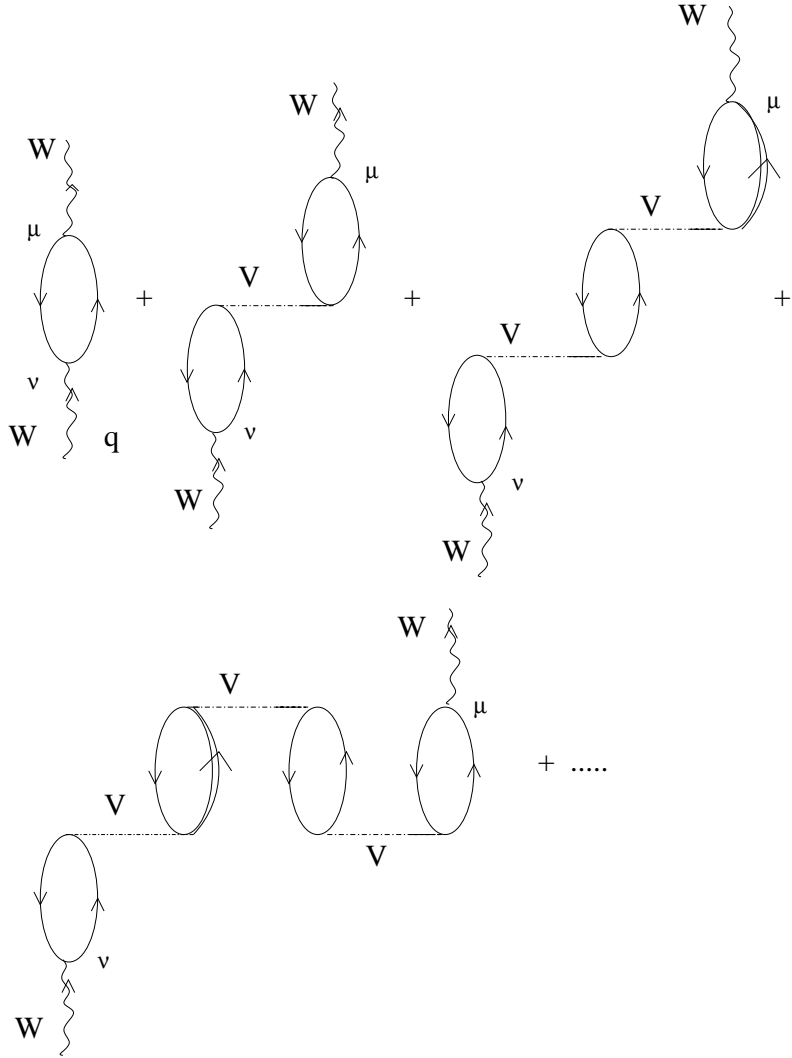}}
\caption{\footnotesize Set of irreducible diagrams responsible for the
  polarization (RPA) effects in the 1p1h contribution to the
  $W-$selfenergy. }\label{fig:fig3}
\end{figure}

When the electroweak interactions take place in nuclei, the strengths
of electroweak couplings may change from their free nucleon values due
to the presence of strongly interacting nucleons~\cite{Ch90}. Indeed,
since the nuclear experiments on $\beta$ decay in the early seventies
\cite{Wi74}, the quenching of axial current is a well established
phenomenon. We follow here the MBF of Ref.~\cite{GNO97}, and take into
account the medium polarization effects in the 1p1h contribution to
the $W-$selfenergy by substituting it
by an RPA response as shown diagrammatically in
Fig.~\ref{fig:fig3}. For that purpose we use an effective ph--ph
interaction of the Landau-Migdal type
\begin{equation}
\begin{array}{ll}
V = & c_{0}\Big\{
f_{0}(\rho)+f_{0}^{\prime}(\rho)\vec{\tau}_{1}\vec{\tau}_{2}+ 
g_{0}(\rho)\vec{\sigma}_{1}\vec{\sigma}_{2}+g_{0}^{\prime}(\rho)
\vec{\sigma}_{1}\vec{\sigma}_{2}
\vec{\tau}_{1}\vec{\tau}_{2}
\Big\}
\end{array}
\end{equation}
where $\vec{\sigma}$ and $\vec{\tau}$ are Pauli matrices acting on the
nucleon spin and isospin spaces, respectively. Note that the above
interaction is of contact type, and therefore in  coordinate
space one has $V(\vec{r}_{1},\vec{r}_{2}) \propto
\delta(\vec{r}_{1}-\vec{r}_{2})$. As mentioned before, the coefficients
were determined in Ref.~\cite{Sp77} from calculations of nuclear
electric and magnetic moments, transition probabilities, and giant
electric and magnetic multipole resonances. They are parameterized as
\begin{equation}
f_{i}(\rho (r))=\frac{\rho (r)}{\rho (0)} f_{i}^{(in)}+
\left[ 1-\frac{\rho (r)}{\rho (0)}\right] f_{i}^{(ex)}
\end{equation}
where
\begin{equation}
\begin{array}{ll}
f_{0}^{(in)}=0.07 & f_{0}^{\prime (ex)}=0.45\\ 
f_{0}^{(ex)}=-2.15 &
f_{0}^{\prime (in)}= 0.33 \\ 
g_{0}^{(in)}=g_{0}^{(ex)}=g_{0}=0.575 ~~~~~~~~~&
g_{0}^{\prime (in)}=g_{0}^{\prime (ex)}=g_{0}^{\prime}=0.725 \\
\end{array}
\end{equation}
and $c_{0}=380\, {\rm MeV fm}^{3}$. In the $S = 1 = T$ channel
($\vec{\sigma} \vec{\sigma} \vec{\tau} \vec{\tau}$ operator) we use an
interaction with explicit $\pi$ (longitudinal) and $\rho$ (transverse)
exchanges, which has been used for the renormalization of the pionic
and pion related channels in different nuclear reactions at
intermediate energies~\cite{GNO97}, \cite{CO92}--\cite{pion}. Thus we replace,
\begin{equation}
c_0g_{0}^{\prime}(\rho) \vec{\sigma}_{1}\vec{\sigma}_{2}
\vec{\tau}_{1}\vec{\tau}_{2} \to \vec{\tau}_{1}\vec{\tau}_{2}
\sum_{i,j=1}^3\sigma_{1}^i \sigma_{2}^j V_{ij}^{\sigma\tau}, \qquad
V_{ij}^{\sigma\tau} = \left (\hat{q}_{i}\hat{q}_{j}V_{l}(q)
+({\delta}_{ij}- \hat{q}_{i}\hat{q}_{j})V_{t}(q)\right) \label{eq:st1}
\end{equation}
with $\hat{q} = \vec{q}/|\vec{q}\,|$ and the strengths of the ph-ph
interaction in the longitudinal and transverse channel are given by
\begin{eqnarray}
V_l(q^0,\vec{q}) &=& \frac{f^2}{m^2_\pi}\left
\{\left(\frac{\Lambda_\pi^2-m_\pi^2}{\Lambda_\pi^2-q^2 }\right)^2
\frac{\vec{q}{\,^2}}{q^2-m_\pi^2} + g^\prime_l(q)\right \}, \qquad
\frac{f^2}{4\pi}=0.08,~~\Lambda_\pi=1200~{\rm MeV} \nonumber \\
V_t(q^0,\vec{q}) &=& \frac{f^2}{m^2_\pi}\left
\{   C_\rho \left (\frac{\Lambda_\rho^2-m_\rho^2}{\Lambda_\rho^2-q^2 }\right)^2
\frac{\vec{q}{\,^2}}{q^2-m_\rho^2} + g^\prime_t(q)\right \}, \qquad
C_\rho=2,~~\Lambda_\rho=2500~{\rm MeV},~~m_\rho=770~{\rm MeV} \label{eq:st2}
\end{eqnarray}
The SRC functions $g^\prime_l$ and $g^\prime_t$ have a smooth
$q-$dependence~\cite{OTW82,Ga88}, which we will not consider
here\footnote{This is justified because taking into account the
$q-$dependence leads to minor changes for low and intermediate
energies and momenta, where this effective ph-ph interaction should be
used.}, and thus we will take
$g^\prime_l(q)=g^\prime_t(q)=g^\prime=0.63$ as it was done in the
study of inclusive nuclear electron scattering carried out in
Ref.~\cite{GNO97}, and also in some of the works of
Ref.~\cite{pion}. Note that, $c_0 g^\prime_0$ and $g^\prime
f^2/m_\pi^2$ differ from each other in less than 10\%.

We also include $\Delta(1232)$ degrees of freedom in the nuclear
medium which, given the spin-isospin quantum numbers of the $\Delta$
resonance, only modify the vector-isovector ($S = 1 = T$) channel of
the RPA response function. The ph--$\Delta$h and $\Delta$h--$\Delta$h
effective interactions are obtained from Eqs.~(\ref{eq:st1})
and~(\ref{eq:st2}) by replacing $\vec{\sigma} \to \vec{S} $,
$\vec{\tau} \to \vec{T} $, where $\vec{S},\vec{T} $ are the spin,
isospin $N\Delta$ transition operators~\cite{OTW82} and $f\to
f^*=2.13~f$, for any $\Delta$ which replaces a nucleon.

Thus, the $V$ lines in Fig.~\ref{fig:fig3} stand for the
effective ph($\Delta$h)-ph($\Delta$h)  interaction described so
far. Given the isospin structure of the $W^{\pm}NN$ coupling, the
isoscalar terms ($f_0$ and $g_0$) of the effective interaction can not
contribute to the RPA response function. We should stress
that this effective interaction is non-relativistic, and then for
consistency we will  neglect terms of order ${\cal O}(p^2/M^2)$ when 
summing up the RPA series. 

To start with, let us examine how the axial vector term ($G_A \gamma^\mu
\gamma_5 \tau_+/2$, where $\tau_{\pm} = \tau_x \pm {\rm i} \tau_y$ are the
ladder isospin operators responsible for the $n$ to $p$ and $p$ to $n$
transitions, $\tau^+|n> = 2 |p>$)   of the CC axial current is
renormalized. As mentioned above, we will only compute the higher
density corrections, implicit in the RPA series, to the leading and
next-to-leading orders in the $p/M$ expansion. The nonrelativistic
reduction of the axial vector term in the nucleon current reads 
\begin{equation}
 G_A \overline{u}_{r^\prime}(\vec{p}^{\,\prime}) \frac{\tau_+}{2}
 \gamma^\mu \gamma_5 u_r(p) = 2M G_A   
 \chi^\dagger_{r^\prime}  \left ( -g^{\mu i} \sigma^i + g^{\mu
 0}\frac{\vec{\sigma}\cdot(\vec{p}+\vec{p}^{\,\prime})}{2M} + \cdots
 \right )\frac{\tau_+}{2} \chi_r, \quad i=1,2,3 \label{eq:axnorel}
\end{equation}
with $\vec{p}^{\,\prime}=\vec{p} +\vec{q}$, and $\chi_r$ a
non-relativistic nucleon spin-isospin wave-function. In
Eq.~(\ref{eq:axnorel}) there is a sum on the repeated index $i$ and
the dots stand for corrections\footnote{Note that $q^0/M$
is of the order $|\vec{q}\,|^2/M^2$.} of order ${\cal
O}\left(\vec{p}^{\,2}/M^2,\vec{p}^{\, \prime 2}/M^2,q^0/M\right)$. In
the impulse approximation, this current leads to a CC nucleon
tensor\footnote{Keeping up to next-to-leading terms in the $p/M$
expansion.},
\begin{equation}
A^{\mu\nu}(p,q)|_{{\rm axial~ vector}}^{NR} = 8M^2 \left ({\cal A}_1^{\mu\nu}
+ {\cal A}_2^{\mu\nu} \right ), \quad {\cal A}_1^{\mu\nu} =  G_A^2  g^{\mu
  i}g^{\nu j}\delta^{ij}, \quad {\cal A}_2^{\mu\nu} = - G_A^2  \left (g^{\mu
  i}g^{\nu 0} + g^{\mu 0}g^{\nu i} \right ) \frac{(2\vec{p} +
  \vec{q}\,)^i}{2M}   \label{eq:defa1a2}
\end{equation}
with $i,j=1,2,3$ and there is again a sum for repeated indices. The tensor
$A^{\mu\nu}(p,q)|_{{\rm axial~ vector}}^{NR}$ can be also obtained from
the non-relativistic reduction of $A^{\mu\nu}(p,q)$ in
Eq.~(\ref{eq:nucl}). The ${\cal A}_1$ contribution comes from the
leading operator $-g^{\mu i}\sigma^i \tau_+/2$, and involves the trace
of $G_A^2 \sigma^i \sigma^j $ (1p1h excitation depicted in the first
diagram of Fig.~\ref{fig:fig3}). Let us consider first this simple
operator and only forward propagating (direct term of the Lindhard
function) ph--excitations. Taking into
account the spin structure of this operator the scalar term
$f_{0}^{\prime}$ of the effective interaction does not contribute
either, and thus we are left with the spin-isospin channel of the
effective interaction, $\sum_{ij}V_{ij}^{\sigma\tau} \sigma_{1}^i \sigma_{2}^j
\vec{\tau}_{1}\vec{\tau}_{2}$. Let us now look at the irreducible
diagrams consisting of the excitation of one and two ph states (first
and second diagrams of Fig.~\ref{fig:fig3}). The
contribution of those diagrams to the W-selfenergy is
\begin{eqnarray}
\Pi^{ij}_W &\propto& ~<p| \frac{\tau_+}{2}|n> <n| \frac{\tau_-}{2}|p>
\frac{\overline {U}}{2} Tr(\sigma^i\sigma^j)\nonumber \\ &+& <p|
\frac{\tau_+}{2}|n><n|\vec{\tau}|p><p|\vec{\tau}|n> <n|
\frac{\tau_-}{2}|p> \left (\frac{\overline {U}}{2} \right)^2
\sum_{k,l=1}^3 Tr(\sigma^i\sigma^l)Tr(\sigma^k\sigma^j)V^{\sigma\tau}_{lk} \nonumber \\ &=&
\overline {U}(q,k_F^n,k_F^p) \left ( \delta^{ij} + 2  \overline
{U}(q,k_F^n,k_F^p) V^{ij}_{\sigma\tau} \right ) \label{eq:2ph}
\end{eqnarray}
The excitation of three ph states gives a contribution of
$\overline{U} (2\overline{U})^2\sum_k
V^{ik}_{\sigma\tau}V^{kj}_{\sigma\tau}= \overline{U} (2\overline{U})^2
\left ( \hat{q}_{i}\hat{q}_{j}V_{l}^2 +({\delta}_{ij}-
\hat{q}_{i}\hat{q}_{j})V_{t}^2 \right )$ to $\Pi^{ij}_W$. Thus, the
full sum of multiple ph excitation states, implicit in
Fig.~\ref{fig:fig3}, leads to two independent geometric series, in
the longitudinal and transverse channels, which are taken into account
by the following substitution  in the hadronic tensor ($W^{\mu\nu}$)
\begin{eqnarray}
 \delta^{ij} 8 M^2 G_A^2 {\rm Im} \overline {U}(q,k_F^n,k_F^p) &\to& 8
 M^2 G_A^2
 {\rm Im}\left \{ \overline
 {U}(q,k_F^n,k_F^p) \left (
 \frac{\hat{q}^i\hat{q}^j}{1-2\overline {U}(q,k_F^n,k_F^p)V_l(q)} + 
\frac{\delta^{ij}-\hat{q}^i\hat{q}^j}{1-2\overline {U}(q,k_F^n,k_F^p)
V_t(q)}\right) \right \} \nonumber \\
&=& \hspace{-0.2cm}8 M^2 G_A^2
 {\rm Im} \overline
 {U}(q,k_F^n,k_F^p) \left (
 \frac{\hat{q}^i\hat{q}^j}{|1-2\overline {U}(q,k_F^n,k_F^p)V_l(q)|^2} + 
\frac{\delta^{ij}-\hat{q}^i\hat{q}^j}{|1-2\overline {U}(q,k_F^n,k_F^p)
V_t(q)|^2}\right) \label{eq:axrpa}
\end{eqnarray}
The factor 2 in the denominator above and that in Eq.~(\ref{eq:2ph})
comes from the isospin dependence,
$\vec{\tau}_{1}\cdot\vec{\tau}_{2}$, of the effective ph--ph
interaction. Taking account of $\Delta$h and backward (crossed term of
the Lindhard function) propagating ph excitations (see
Fig.~\ref{fig:fig3}), not accounted for by $\overline{U}$ is readily
done by substituting $2\overline{U}$ in the denominator by
$U(q,k_F)=U_N+U_\Delta$, the Lindhard function of Ref.~\cite{Ga88},
which for simplicity we evaluate\footnote{The functions $U_N$ and
$U_\Delta$ are defined in Eqs.(2.9) and (3.4) of
Ref.~\protect\cite{Ga88}, respectively. Besides, note that in a
symmetric nuclear medium $U_N= 2\overline{U}+ $ backward propagating
ph excitation. For positive values of $q^0$ the backward propagating
ph excitation has no imaginary part, and for QE kinematics $U_\Delta$
is also real.} in an isospin symmetric nuclear medium of density
$\rho$. The different couplings for $N$ and $\Delta$ are incorporated
in $U_N$ and $U_\Delta$ and then the same interaction strengths $V_l$
and $V_t$ are used for ph and $\Delta $h
excitations~\cite{OTW82,pion}. Taking $\vec{q}$ in the $z$ direction, 
Eq.~(\ref{eq:axrpa}) implies that the axial vector
contribution to the transverse ($xx,yy$) and longitudinal ($zz$)
components of the hadronic tensor get renormalized by different
factors $1/|1-U(q,k_F)V_t(q)|^2$ versus $1/|1-U(q,k_F)V_l(q)|^2$.

Let us pay now attention to the term ${\cal A}_2$ in
Eq.~(\ref{eq:defa1a2}), it comes from the interference between the 
$-g^{\mu i}\sigma^i \tau_+/2$ and $ \tau_+ g^{\mu
 0}\left (\vec{\sigma}\cdot(\vec{p}+\vec{p}^{\,\prime})\right)/4M$
operators in Eq.~(\ref{eq:axnorel}). The consideration of the full RPA
series leads now to the substitution\footnote{To evaluate the
  longitudinal contribution we use $2p_z+q_z= (2\vec{p} +
  \vec{q})\cdot \vec{q} / |\vec{q}\,| = q^0\left
  (2E(\vec{p}\,)+q^0\right)/ |\vec{q}\,| = 
  2Mq^0 / |\vec{q}\,|+{\cal O}(\vec{p}^{\,2}/M^2,q^0/M)$. Besides,
  the transverse part of the effective interaction does not contribute
  since $\left (\delta_{zk}-\hat{q}_z\hat{q}_k \right ) (2\vec{p} +
  \vec{q}\,)_k = 0$.} 
\begin{equation}
8 M^2 \frac{q^0}{|\vec{q}\,|} G^2_A {\rm
  Im}\overline{U}(q,k_F^n,k_F^p) \to 8 M^2 \frac{q^0}{|\vec{q}\,|}
  G^2_A    \frac{{\rm
  Im}\overline{U}(q,k_F^n,k_F^p)}{|1-U(q,k_F)V_l(q)|^2} \label{eq:a0zrpa}
\end{equation}
in the $0z$ and $z0$ components of the hadronic tensor
$W^{\mu\nu}$. 

Keeping track of the responsible operators, we have examined and
renormalized all different contributions to the CC nucleon tensor
$A^{\mu\nu}$, by summing up the RPA series depicted in
Fig.~\ref{fig:fig3}. The $00,0z,zz,xx$ and $xy$ components of the RPA
renormalized CC nucleon tensor\footnote{These are the needed
components to compute the hadronic tensor $W^{\mu\nu}$, when $\vec{q}$
is taken in
the $z$ direction.} can be found in Sect.~\ref{sec:rpaamunu} of the
Appendix. As mentioned above, since the ph($\Delta$h)-ph($\Delta$h)
effective interaction is non-relativistic, we have computed
polarization effects only for the leading and next-to-leading terms in
the $p/M$ expansion. Thus, order ${\cal
O}\left(k_F\vec{p}^{\,2}/M^2,k_F\vec{p}^{\, \prime
2}/M^2,k_Fq^0/M\right)$ has been neglected in the formulae of the
Sect.~\ref{sec:rpaamunu} of the Appendix. We have made an exception
to the above rule, and since $\mu_V$ could be relatively large, we
have taken $\mu_V F_2^V|\vec{q}\,|/M$ to be of order ${\cal O}(0)$ in
the $p/M$ expansion. Finally, we should stress that the
scalar--isovector term of the effective interaction ($f^\prime$)
cannot produce $\Delta$h excitations and therefore, when this term is
involved in the RPA renormalization, only the nucleon Lindhard
function ($U_N$) appears (see coefficient ${\bf C_N}$ in
Eq.~(\ref{eq:coeffs})).

To finish this subsection we will discuss the  differences
between the medium polarization scheme presented here and that
undertaken in Refs.~\cite{Ch90,Si92}. There is an obvious
difference, since in these latter references the scalar-isovector term
($f^\prime$) of the ph--ph  effective interaction
was not taken into account. In addition, there are some differences
concerning the tensorial treatment of the RPA response function.
In the framework presented in this work we firstly evaluate the 1p1h
hadronic tensor and all sort of polarization (RPA) corrections to the
different components of this tensor. In a second step we contract it
with the leptonic tensor and obtain the differential cross
section\footnote{Note that the differential cross section is determined by 
 the $00$, $0z$, $zz$, $xx$ and $xy$ components of
$W^{\mu\nu}$ through their relation to the $W_{i=1,\cdots, 5}$
structure functions. See Eqs.~(\protect\ref{eq:ws})
and~(\protect\ref{eq:cross}).}. The RPA corrections do not depend only on the
different terms of the nucleon currents, but also on the particular
component of the hadronic tensor which is being renormalized. Thus, as
it is obvious, the RPA corrections, in general, are different for
each of the terms ($(F_1^V)^2, (F_2^V)^2, F_1^VF_2^V,
G_A^2,G_P^2,G_AG_P,F_1^VG_A$ and $F_2^VG_A$, with
$G_P=2MG_A/(m^2_\pi-q^2) $ ) appearing in the CC nucleon tensor. Besides,
for a fixed term, the polarization effects do also depend on the
tensor component. Indeed, we have already mentioned this fact in the
discussion of Eq.~(\ref{eq:axrpa}), where we saw that the axial vector
contribution to the transverse ($xx$,$yy$) and the longitudinal ($zz$)
components of the hadronic tensor get  renormalized by different
factors.

In the works of Refs.~\cite{Ch90, Si92} the 1p1h hadronic tensor,
without polarization effects included, is first contracted with the
leptonic one. This contraction is denoted as, up to global kinematical
factors, $\overline{\sum}\sum |T|^2$ in those references. In a second
step, the authors of Refs.~\cite{Ch90,Si92} study the medium
polarization corrections to $\overline{\sum}\sum |T|^2$. They find out
different medium corrections for each of the terms of  the CC
nucleon tensor ($(F_1^V)^2,\cdots , F_2^VG_A$), as we do. However, for
a fixed term, they cannot independently study the effect of the RPA
re-summation in each of the different tensor components, since they
are not dealing with the hadronic tensor itself, but with the
contraction of it with the leptonic one.  As a matter of example, to
account for the RPA corrections to the axial vector--axial vector
term, the following substitution is given in Refs.~\cite{Ch90,Si92}
\begin{equation}
G_A^2 \to G_A^2 \left (\frac{2}{3|1-U(q,k_F)V_t(q)|^2} +
\frac{1}{3|1-U(q,k_F)V_l(q)|^2}    \right )  \label{eq:eul}
\end{equation}
The above substitution can be recovered from Eq.~(\ref{eq:axrpa}) by
contracting this latter equation with $\delta_{ij}$, and replacing
$2\overline{U} \to U $. Thus, Eq.~(\ref{eq:eul}) is strictly correct,
neglecting terms\footnote{ Medium renormalization effects are taken
into account in these terms by means of the substitution of
Eq.~(\ref{eq:a0zrpa}).} of order $p/M$, only for the contribution to
$\overline{\sum}\sum |T|^2$ obtained from the contraction of the
hadronic tensor with the $g_{\mu\nu}$ term\footnote{Note that the
axial vector--axial vector contribution to $W^{00}$ is order ${\cal
O}(\vec{p}^{\,2}/M^2)$.} of the leptonic one. The prescription of
Eq.~(\ref{eq:eul}) is not correct for those contributions to
$\overline{\sum}\sum |T|^2$ arising from the contraction of the
$k^\prime_\mu k_\sigma +k^\prime_\sigma k_\mu$ terms of the leptonic
tensor with the axial vector--axial vector contribution of
$W^{\mu\nu}$. Note however that neglecting the bound muon three
momentum\footnote{This is an accurate approximation and we will also
make use of it in Sect.~\ref{sec:imc}.} and up to terms of order $p/M$,
Eq.~(\ref{eq:eul}) is correct for the study of inclusive muon capture
in nuclei, where it was first used by the authors of
Refs.~\cite{Ch90,Si92}, and it is also reasonable for 
neutrino--nucleus reactions at low energies, where the RPA effects are
more important.

\subsection{Correct Energy Balance and Coulomb Distortion Effects} 
\label{sec:qvalue}
 To ensure the correct energy balance in the reaction~(\ref{eq:reac})
for finite nuclei, the energy conserving $\delta$ function in
Eq.~(\ref{eq:res}) has to be modified~\cite{Ch90,Si92}. The energies
$E(\vec{p})$ and $E(\vec{p}+\vec{q}~)$ in the argument of the $\delta$
function  refer to the LFG of the nucleons in the initial and final
nucleus. In the Fermi sea there is no energy gap for the transition from
the occupied to the unoccupied states and hence ph excitations can be
produced with a small energy, $Q^{\rm LFG}(r)= E_F^p(r)-E_F^n(r)$
. However, in actual nuclei there is a minimum excitation energy,
$Q=M(A_{Z+1})-M(A_Z)$, needed for the transition to the ground state
of the final nucleus. For instance , this $Q$ value is
16.827 MeV for the transition $^{12}{\rm C}_{gs} \to ^{12}{\rm
N}_{gs}$ and the consideration of this energy gap is essential to
obtain reasonable cross sections for low-energy neutrinos. We have
taken it into account by replacing
\begin{equation}
q^0 \to q^0
-(Q-Q^{\rm LFG}(r)) \label{eq:qvalue}
\end{equation} 
in the $\delta-$function of the right hand side of Eq.~(\ref{eq:res}).

The second effect which we want to address here is due to the fact
that the charged lepton produced in the reaction of
Eq.~(\ref{eq:reac}) is  moving in the Coulomb field of the nucleus
described by a charge distribution $\rho_{ch}(r)$. In our scheme, we
implement the corrections due to this effect following the
semiclassical approximation used in Ref.~\cite{Si92}. Thus, we include
a selfenergy (Coulomb potential) in the intermediate lepton
propagator of the neutrino selfenergy depicted in
Fig.~\ref{fig:fig1}. We approximate this selfenergy inside the LFG
by
\begin{equation}
\Sigma_C = 2k^{\prime 0} V_C(r), \quad V_C(r) = - 4\pi \alpha \left (
\frac{1}{r} \int_0^r dr^\prime r^{\prime 2} \rho_{ch}(r^\prime) 
+ \int_r^{+\infty} dr^\prime r^\prime  \rho_{ch}(r^\prime)\right )
\end{equation}
with $\alpha=1/137.036$ and the charge distribution, $\rho_{ch}$, normalized to
$Z$. The evaluation of the imaginary part of the $\nu$ self-energy in the
medium requires to put  the intermediate lepton
propagator on the mass shell. Following the Cutkosky's rules, and 
neglecting quadratic corrections in $V_C$, we find
\begin{equation}
\frac{1}{k^{\prime\,2} -m^2_l-2 k^{\prime 0}V_C(r)+ {\rm i}\epsilon} 
\to -{\rm i}\pi\frac{\delta(k^{\prime 0}
-E^\prime_l)}{k^{\prime 0}}
  \Theta(\hat{E}^\prime_l(r)-m_l)  
\end{equation}
where $E^\prime_l$ is the asymptotical outgoing lepton energy in regions
  where the Coulomb potential can be neglected, and the local
outgoing lepton energy, $\hat{E}^\prime_l(r)$, is defined by 
 energy conservation 
\begin{equation}
\hat{E}^\prime_l(r)+V_C(r)= 
\sqrt{m_l^2+\vec{{\cal K}}^{\,\prime 2}(r)}+V_C(r)= E^\prime_l 
\end{equation}
Because of the Coulomb potential, the outgoing lepton three momentum,
$\vec{{\cal K}}^{\,\prime}$, is not longer conserved, and it becomes a
function of $r$, taking its asymptotical value, $\vec{k}^{\,\prime}$,
at large distances.  Therefore, $\vec{q}$ should also be replaced by a
local function: $\vec{q}^{\,\,\prime}(r)=\vec{k}-\vec{{\cal
K}}^{\,\prime}(r)$.  Furthermore, from the $d^3k^\prime$ integration in
Eq.~(\ref{eq:ims}), and considering now the locality of the three
momentum, we get from phase space a correction factor $|\vec{{\cal
K}}^{\,\prime}(r)| \hat{E}^\prime_l(r) /
|\vec{k}^{\,\prime}|E^\prime_l $.  This way of taking into account the
Coulomb effects has clear resemblances with what is called ``modified
effective momentum approximation'' in Ref.~\cite{En98}. The use of a
plane-wave approximation in the interaction region is equivalent to
the assumption that the Coulomb potential does not change the
direction of the particles when they leave the nucleus. It should
therefore not strongly alter an outgoing negatively charged lepton
wave packet, which asymptotically is spherical, after it leaves the
nucleus except by slowing it down and thereby changing the average
radial wavelength and amplitude as the wave moves to larger $r$.  As
it is shown in~\cite{En98}, for total cross sections this procedure
works very accurately for muons down to low energies.  For low- energy
electrons and positrons it is less accurate, and the use of the Fermi
function $F(Z,E^\prime_l)$ (\cite{Be82}) is widely accepted in the
literature. Anyway, Coulomb effects are small and they become
relatively sizeable only for neutrino induced reactions near threshold
and/or for heavy nuclei.

To summarize the results of Subsections~\ref{sec:rpa}
and~\ref{sec:qvalue}, our final expression  for the hadronic tensor is
given by
\begin{eqnarray}
W^{\mu\nu}(q^0,\vec{q}\,) &=& - \frac{\cos^2\theta_C}{2M^2}
\int_0^\infty dr r^2 \frac{|\vec{{\cal K}}^{\,\prime}(r)| \hat{E}^\prime_l(r)}{
|\vec{k}^{\,\prime}|E^\prime_l} \Theta(\hat{E}^\prime_l(r)-m_l) 
\Big \{ 2 \Theta(q^{\prime 0}) \int
\frac{d^3p}{(2\pi)^3}\frac{M}{E(\vec{p})}
\frac{M}{E(\vec{p}+\vec{q}^{\,\,\prime})} \times
 \nonumber\\
&& \Theta(k_F^n(r)-|\vec{p}~|)
 \Theta(|\vec{p}+\vec{q}^{\,\,\prime}\,|-k_F^p(r))  
(-\pi)\delta\left(q^{\prime\, 0} + E(\vec{p})
-E(\vec{p}+\vec{q}^{\,\,\prime} \,)\right )
A^{\nu\mu}_{\rm RPA}(p,q^\prime)|_{p^0=E(\vec{p})~} \Big \}
\end{eqnarray}
with $q^{\prime 0}= q^0 -(Q-Q^{\rm LFG}(r))$,
$\vec{q}^{\,\,\prime}(r)=\vec{k}-\vec{{\cal K}}^{\,\prime}(r)$,  and
$A^{\mu\nu}_{\rm RPA}$ given in Sect.~\ref{sec:rpaamunu} of the
Appendix.

\subsection{FSI Effects} 
\label{sec:fsi}

Once a ph excitation is produced by the virtual $W-$boson, the
outgoing nucleon can collide many times, thus inducing the emission of
other nucleons. The result of it is a  quenching of the
QE peak respect to the simple ph excitation calculation and a
spreading of the strength, or widening of the
peak. The integrated strength over energies is not much affected
though. A distorted wave approximation with an optical (complex)
nucleon nucleus potential would remove all these events. However, if
we want to evaluate the inclusive $(\nu_l,l^-)$ cross section these
events should be kept and one must sum over all open final state
channels. 

\begin{figure}[t]
\centerline{\includegraphics[height=6cm]{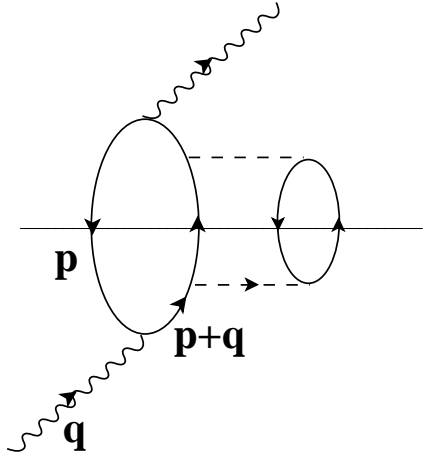}}
\caption{\footnotesize $W^+-$selfenergy diagram obtained from the
  first diagram depicted in Fig.~\protect\ref{fig:fig2} by dressing up
  the nucleon pro\-pa\-ga\-tor of the particle state in the ph
  excitation. }\label{fig:burbuja}
\end{figure}

In our MBF we will account for the FSI by using
nucleon propagators properly dressed with a realistic selfenergy in
the medium, which depends explicitly on the energy and the
momentum~\cite{FO92}. This selfenergy leads to nucleon spectral
functions in good agreement with  accurate  microscopic
approaches like the ones of Refs.~\cite{Ra89,Mu95}. The selfenergy of
Ref.~\cite{FO92} has a proper energy--momentum dependence plus an
imaginary part from the coupling to the 2p2h components, which is
equivalent to the use of correlated wave functions, evaluated from
realistic $NN$ forces and incorporating the effects of the nucleon
force in the nucleon pairs. Thus, we consider the many body diagram
depicted in Fig.~\ref{fig:burbuja} (there the dashed lines stand
 for an $NN$ interaction inside of the nuclear medium~\cite{OTW82,FO92}). 
However, a word of caution is in order since
the imaginary part of this
diagram presents a divergency. The reason is that when placing the 2p2h
excitation on the mass shell through Cutkosky rules, we still have the square
of the nucleon propagator with momentum $p + q$ in the figure. This
propagator can be placed on shell for virtual $W-$bosons 
 and we get a divergence.
\begin{figure}[b]
\centerline{\includegraphics[height=6cm]{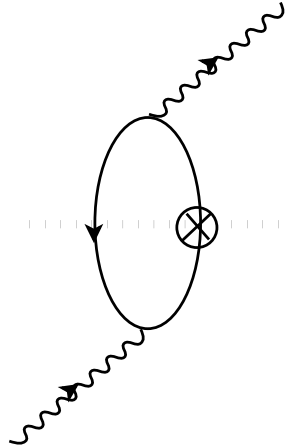}}
\caption{\footnotesize Insertion of the nucleon
selfenergy on the nucleon line of the particle state. }\label{fig:burbuja2}
\end{figure}
The divergence is not spurious, in the sense that its meaning is the
probability per unit time of absorbing a virtual $W^+$ by one nucleon
times the probability of collision of the final nucleon with other
nucleons in the infinite Fermi sea in the lifetime of this
nucleon. Since this nucleon is real, its lifetime is infinite and thus
the probability is infinite, as well. The problem is physically solved~\cite{Sa88}
by recalling that the nucleon in the Fermi sea has a selfenergy with
an imaginary part which gives it a finite lifetime (for
collisions). This is taken into account by iterating, in the Dyson
equation sense, the nucleon selfenergy insertion of
Fig.~\ref{fig:burbuja2} in the nucleon line, hence substituting the
particle nucleon propagator, $G(p;\rho)$, in Eq.~(\ref{eq:Gp}) by a
renormalized nucleon propagator, $G_{\rm FSI}(p;\rho)$, including the
nucleon selfenergy in the medium, $\Sigma (p^0 , \vec{p}\,; \rho)$,
\begin{equation}
 G_{\rm FSI}(p;  \rho)= \frac{1}{p^{0}-{\bar E}(\vec{p}\,)-\Sigma(p^{0},
  \vec{p}\,;\rho)} \label{eq:gfsi}
\end{equation}
with ${\bar E}(\vec{p}\,) = M + \vec{p}^{\,2}/2M$. 
As mentioned above, we use here the nucleon
selfenergy model developed in Ref.~\cite{FO92}, which led to
excellent results in the study of inclusive electron scattering from
nuclei~\cite{GNO97}. Since the model of Ref.~\cite{FO92} is not
Lorentz relativistic and it also considers an isospin symmetric
nuclear medium, we will only discuss the FSI effects for nuclei with
approximately equal number of protons and neutrons, and using 
non-relativistic kinematics for the nucleons (see
Sect.~\ref{sec:nonrel} of the Appendix). Thus, we have obtained
Eq.~(\ref{eq:gfsi}) from the non-relativistic reduction of
$G(p;\rho)$, in Eq.~(\ref{eq:Gp}), by including the nucleon selfenergy.

Alternatively to the nucleon selfenergy language,
one can use the spectral function representation 
\begin{equation}
\label{eq:spsh}
G_{\rm FSI} (p;\rho) = \int_{- \infty}^\mu  \frac{S_h (\omega, 
\vec{p}\,;\rho)}{p^0  -\omega - i \epsilon} d \omega + 
\int_{\mu}^\infty \frac{S_p (\omega, \vec{p}\,;\rho)}{p^0 -\omega + i
\epsilon} d \omega \label{eq:gfsi2}
\end{equation}
where $S_h, S_p$ are the hole and particle spectral functions related to
nucleon selfenergy $\Sigma$ by means of
\begin{eqnarray}
S_{p,h}(\omega,\vec{p}\,;\rho) &=& \mp\frac{1}{\pi}\frac{{\rm
Im}\Sigma(\omega,\vec{p}\,;\rho)}{\Big[\omega-{\bar E}(\vec{p}\,)-{\rm
Re}\Sigma(\omega,\vec{p}\,;\rho)  \Big]^2+\Big[{\rm
Im}\Sigma(\omega,\vec{p}\,;\rho)\Big]^2}
\end{eqnarray}
with $\omega\ge \mu$ or $\omega\le \mu$  for $S_p$ and $S_h$,
respectively.  The chemical potential $\mu$ is determined by
\begin{equation}
\mu = M + \frac{k_F^2}{2M} + {\rm Re}\Sigma(\mu, k_F)
\end{equation}
By means of Eq.~(\ref{eq:gfsi2})
 we can write the ph propagator or new Lindhard
function incorporating the effects of the nucleon selfenergy in
the medium, and we have for its imaginary part (for positive values of $q^0$)
\begin{equation}
{\rm Im} \overline{U}_{FSI}(q;k_F) = 
-\frac{\Theta(q^0)}{4\pi^2}\int d^3p 
 \int_{\mu-q^0}^\mu d\omega S_h(\omega,\vec{p}\,;\rho)
S_p(q^0+\omega,\vec{p}+\vec{q}\,;\rho) 
\end{equation}
Comparing the above expression with that of the ordinary imaginary
part of the non-relativistic Lindhard function, Eq.~(\ref{eq:imu}), one
realizes that to account for FSI effects in an isospin symmetric
nuclear medium of density $\rho$ we should make the following
substitution
\begin{eqnarray}
& & 2 \Theta(q^0) \int \frac{d^3p}{(2\pi)^3}
\Theta(k_F^n(r)-|\vec{p}~|) \Theta(|\vec{p}+\vec{q}~|-k_F^p(r))
   (-\pi)\delta(q^0 +
{\bar E}(\vec{p}) -{\bar E}(\vec{p}+\vec{q}~))
A^{\nu\mu}(p,q)|_{p^0={\bar E}(\vec{p})~} \nonumber \\
&\rightarrow&  -\frac{\Theta(q^0)}{4\pi^2}\int d^3p 
 \int_{\mu-q^0}^\mu d\omega S_h(\omega,\vec{p}\,;\rho)
S_p(q^0+\omega,\vec{p}+\vec{q}\,;\rho) 
A^{\nu\mu}(p,q)|_{p^0={\bar E}(\vec{p})~} \label{eq:fsi}
\end{eqnarray}
in the expression of the hadronic tensor (Eq.~(\ref{eq:res})). The
$d^3p$ integrations have to be  done numerically. Indeed, 
the integrations are not trivial from the
computational point of view, since in some regions the spectral functions
behave like delta functions. We use the spectral functions calculated
in Ref.~\cite{FO92}, but since the imaginary part of the nucleon
selfenergy for the hole states is much smaller than that of the
particle states at intermediate nuclear excitation energies, we make
the approximation of setting to zero Im$\Sigma$ for the hole
states. This was found to be a good approximation in
\cite{Ci90}. Thus, we take
\begin{equation}
S_{h}(\omega,\vec{p}\,;\rho)=\delta(\omega-\hat{E}(\vec{p}\,))
\Theta(\mu-\hat{E}(p))
\end{equation}
where $\hat{E} (p)$ is the energy associated to a momentum $\vec{p}$
obtained self consistently by means of the equation
\begin{equation}
\hat{E}(\vec{p}\,)={\bar E}(\vec{p}\,) + {\rm Re} \Sigma (\hat{E}(\vec{p}\,),
    \vec{p}\,;\rho)
\end{equation}
It must be stressed that it is important to keep the real part of 
$\Sigma$ in the hole states when renormalizing the particle states because
there are terms in the nucleon  selfenergy largely independent of the 
momentum and which cancel in the ph propagator, where the two 
selfenergies subtract.

On top of the FSI corrections examined here, one should also take into
account the nuclear corrections studied previously in
Subsections~\ref{sec:rpa} and~\ref{sec:qvalue}.

\section{CC  Antineutrino Induced Nuclear Reactions}\label{sec:anti}

The cross section for  the antineutrino induced nuclear reaction
\begin{equation}
{\bar \nu}_l (k) +\, A_Z \to l^+ (k^\prime) + X  \label{eq:anti}
\end{equation}
is easily obtained from the expressions given in Sects.~\ref{sec:neu}
and~\ref{sec:self} with the followings modifications:
\begin{itemize}

\item Changing  the sign of  the parity-violating 
  terms, proportional to $W_3$, in the differential cross
  section, Eq.~(\ref{eq:cross}).

\item Replacing the $W^+ $selfenergy in the medium, $\Pi_W^{\mu\nu}$,
  by that of the $W^-$ boson ($\overline{\Pi}_W^{\mu\nu}$). This is
  achieved by exchanging the role of protons and
  neutrons in all formulae,
$\overline{\Pi}_W^{\mu\nu} \left(\rho_p(r),\rho_n(r)\right ) =
  \Pi_W^{\mu\nu}\left(\rho_n(r),\rho_p(r)\right ) $. 

\item Changing  the sign of $V_C$, which turns out to be repulsive for 
positive charged outgoing leptons.

\item Correcting the LFG energy balance with the difference
  $\overline{Q} - \overline{Q}^{\rm LFG}(r)$, with  
  $\overline{Q}=M(A_{Z-1})-M(A_Z)$ and $\overline{Q}^{\rm LFG}(r) = 
  E_F^n(r)-E_F^p(r)$.

\end{itemize}

\section{Inclusive Muon Capture in  Nuclei} \label{sec:imc}

In this section we study the $\mu-$atom inclusive decay, it is to say
the reaction
\begin{equation}
\left (A_Z-\mu^- \right )_{\rm bound}^{1s} \to \nu_\mu(k) + X
\end{equation}
It is obvious that the dynamics that governs this process is 
related to that of antineutrino (Eq.~(\ref{eq:anti})) and
neutrino (Eq.~(\ref{eq:reac})) induced nuclear processes, but a
distinctive feature is that the nuclear excitation energies involved
in the $\mu-$atom decay are extremely low (smaller than $\approx$ 20
MeV). In this energy regime one might expect important un-accuracies
in the LFG description of the nucleus. However and due to the inclusive
character of the process, we will see that our MBF leads to reasonable
results, with discrepancies of the order of 10-15\% at most, with RPA
effects as large as a factor of two. We should emphasize that similar
conclusions were achieved in the works of Ref.~\cite{Ch90}, which also
use a LFG picture of the nucleus.

The evaluation of the decay width for finite nuclei proceeds in two
steps. In the first one we evaluate the spin averaged decay width for
a muon at rest\footnote{In what follows, we will neglect the three
momentum of the bound muon.} in a Fermi sea of protons and neutrons
with $N\neq Z$. In this first step, the strong renormalization effects
(RPA) will be also taken into account and thus we will end up with a
decay width, $\hat\Gamma$ which will be a function of the proton and
neutron densities. In the second step, we use the LDA to go to finite
nuclei and evaluate
\begin{equation}
\Gamma = \int d^3 r |\phi_{\rm 1s} (\vec{r}\,)|^2
\hat\Gamma\left(\rho_p(r),\rho_n(r)\right ) \label{eq:imclda}
\end{equation}
where $\phi_{\rm 1s} (\vec{r}\,)$ is the muon wave function in the
$1s$ state from where the capture takes place. It has been obtained by
solving the Schr\"odinger equation with a Coulomb interaction taking
account of the finite size of the nucleus and vacuum
polarization~\protect\cite{Iz80}.  Equation~(\ref{eq:imclda}) amounts to saying
that every bit of the muon, given by the probability $ |\phi_{\rm 1s}
(\vec{r}\,)|^2 d^3 r$, is surrounded by a Fermi sea of 
densities $\rho_p(r),\rho_n(r)$. The LDA assumes a zero range of the
interaction, or equivalently no dependence on $\vec{q}$. The $\vec{q}$
dependence of the interaction is extremely weak for the $\mu-$atom
decay process and, thus the LDA prescription becomes highly
accurate~\cite{Ch90}.
\begin{figure}[b]
\centerline{\includegraphics[height=6cm]{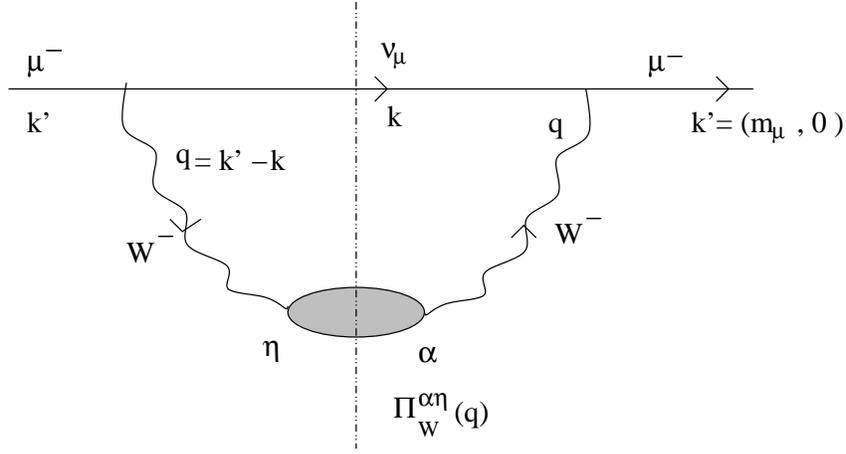}}
\caption{\footnotesize Diagrammatic representation of the muon (at rest)
  selfenergy in nuclear matter. }\label{fig:fig4}
\end{figure}

The spin averaged muon decay width, in an infinite nuclear matter of
densities $\rho_p(r)$ and $\rho_n(r)$, is related to the imaginary
part of the selfenergy (see Fig.~\ref{fig:fig4}),
$\Sigma_\mu^r\left(\rho_p(r),\rho_n(r)\right )$, of a muon at rest and
spin $r$ in the medium by
\begin{equation}
\hat\Gamma\left(\rho_p(r),\rho_n(r)\right ) = -\frac{1}{m_\mu} {\rm
  Im} \Sigma_\mu\left(\rho_p(r),\rho_n(r)\right ), \qquad \Sigma_\mu =
  \frac12 \sum_r \Sigma_\mu^r 
\end{equation}
To evaluate the imaginary part of the selfenergy associated to the diagram of
Fig.~\ref{fig:fig4}, the intermediate states are placed on
shell in the integration over the internal variables. These states are
those crossed by the dash--dotted line in Fig.~\ref{fig:fig4}. The
evaluation of the $\mu-$selfenergy is almost identical to that of a
neutrino in Fig.~\ref{fig:fig1}, and thus we obtain\footnote{There is
a factor $1/2$ of difference coming from the averaged over initial
spins of the muon, besides the $W^--$selfenergy arises since the
negative muon decay process is related to the antineutrino induced
process, and the contribution of parity violating terms flips sign. }
from Eq.~(\ref{eq:ims}),
\begin{equation}
\hat\Gamma\left(\rho_p(r),\rho_n(r)\right ) = -\frac{1}{m_\mu}
\frac{4G}{\sqrt 2 M^2_W}\int \frac{d^3 
  k}{(2\pi)^3 }\frac{\Theta(q^0) }{2|\vec{k}\,|} 
~ {\rm Im}\left\{ \overline{\Pi}^{\mu\eta}_W\left(q;\rho_p(r),\rho_n(r)\right )
 L_{\mu\eta} \right\} 
\end{equation}
with $q^0 = m_\mu-|\vec{k}\,|$ and $|\vec{q}\,|=|\vec{k}\,|$. For
kinematical reasons, only the QE part of the $W^--$selfenergy will
contribute to the muon decay width and thus we find
\begin{eqnarray}
\hat\Gamma\left(\rho_p(r),\rho_n(r)\right ) &=&
 \frac{G^2\cos^2\theta_C}{m_\mu} \int \frac{d^3 k}{(2\pi)^3
 }\frac{1}{2|\vec{k}\,|}  L_{\mu\eta}  {\cal T}^{\mu\eta}
 \left(q;\rho_p,\rho_n\right ) \nonumber \\
&=& \frac{G^2\cos^2\theta_C}{2\pi^2} \int_0^{+\infty}  \vec{k}^{\,2} 
 \left ( -t_1+  \frac{t_2}{2} + |\vec{k}\,| t_3 +
 \frac{m_\mu^2}{2}t_4 + m_\mu t_5
\right ) d|\vec{k}\,| \label{eq:imc2} 
\end{eqnarray}
where the tensor ${\cal T}^{\mu\nu}$ is defined as
\begin{eqnarray}
{\cal T}^{\mu\nu}\left(q;\rho_p,\rho_n\right )  &=&
-\frac{1}{4M^2}\Big \{2 \Theta(q^0) 
\int \frac{d^3p}{(2\pi)^3}\frac{M}{E(\vec{p})}
\frac{M}{E(\vec{p}+\vec{q})} \Theta(k_F^p(r)-|\vec{p}~|)
\Theta(|\vec{p}+\vec{q}~|-k_F^n(r)) \nonumber\\ &\times&
(-\pi)\delta(q^0 + E(\vec{p}) -E(\vec{p}+\vec{q}~)) A^{\mu\nu}_{\rm
RPA}(p,q)|_{p^0=E(\vec{p})~} \Big \} \nonumber \\
&&\nonumber \\
&\equiv& t_1 g^{\mu\nu} + t_2  l^\mu l^\nu + {\rm i} t_3
\epsilon^{\mu\nu\alpha\beta}l_\alpha q_\beta + t_4 q^\mu q^\nu 
 + t_5 \left (l^\mu q^\nu + l^\nu q^\mu \right ), \qquad 
{\rm with}~ l^\mu = (1,\vec{0}). \label{eq:imc}
\end{eqnarray}
The similitude of the above equation with  Eq.~(\ref{eq:res}) is
clear. As in this latter case, the $d^3p$ integrations in
Eq.~(\ref{eq:imc}) can be done analytically (see Sect.~\ref{sec:rel}
of the Appendix) and all of them are determined by the imaginary part
of the relativistic isospin asymmetric Lindhard function, $\overline
{U}_R(q,k_F^n,k_F^p)$. For a non-relativistic Fermi
gas, the decay width is easily obtained from Eq.~(\ref{eq:imc}) 
by replacing the factors $M/E(\vec{p})$ and
$M/E(\vec{p}+\vec{q})$  by one. Analytical
expressions can be now found in the Sect.~\ref{sec:nonrel} of the
Appendix. FSI effects can be also taken into account by performing the
substitution of Eq.~(\ref{eq:fsi}).

Thus, both the muon decay process in the medium and the electroweak
inclusive nuclear reactions $\nu_l (k) +\, A_Z \to l^- (k^\prime) + X
$ in the QE regime are sensitive to the same physical features,
$W^{\pm}pn-$vertex, and RPA and FSI effects. However, in the muon-atom
decay only very small nuclear excitation energies are explored, 0--25
MeV, while in the latter processes higher nuclear excitation energies
can be tested by varying the incoming neutrino momentum.

The $1s$ muon binding energy, $B_\mu^{1s}>0$, can be taken
into account, by replacing $m_\mu \to \hat m_\mu= m_\mu - B_\mu^{1s}$. This
replacement leads to extremely small (significant) changes for light
(heavy) nuclei, where $B^{1s}_\mu$ is of the order of 0.1 MeV (10 MeV),
see Table~\ref{tab:dens}.

Finally, the correct energy balance in the decay can be enforced in
the LFG by replacing 
\begin{equation}
q^0 \to q^0- \left (\overline{Q} - \overline{Q}^{\rm LFG}(r) \right ) =
\hat m_\mu-|\vec{k}\,| - \left (\overline{Q}- \overline{Q}^{\rm LFG}(r) \right )
\end{equation}
in Eqs.~(\ref{eq:imc2}) and~(\ref{eq:imc}). 

\section{Results} \label{sec:res}

Firstly, we compile in Table~\ref{tab:dens} the input used for the
different nuclei studied in this work. Nuclear masses and charge
densities are taken from Refs.~\cite{Fi96} and~\cite{Ja74},
respectively. For each nucleus, we take the neutron matter density
approximately equal (but normalized to $N$) to the charge density, though
we consider small changes, inspired by Hartree-Fock calculations with
the density-matrix expansion~\cite{Ne75} and corroborated by pionic
atom data~\cite{GNO92}.   However charge (neutron) matter
densities do not correspond to proton (neutron) point--like densities
because of the finite size of the nucleon. This is taken  into
account by following the procedure outlined in Section 2 of
Ref.~\cite{GNO92} (see Eqs. (12)-(14) of this reference).
\begin{table}
 \begin{center} \begin{tabular}{llll|rr|c}\hline\tstrut 
 Nucleus & $R_{p}$ [fm]& $R_{n} $[fm] & $a$ [fm]$^*$ &$Q$ [MeV]&
 $\overline{Q}$ [MeV] &$B_\mu^{1s}$  [MeV]  \\\hline \tstrut
 $^{12}$C &1.692 & 1.692 & 1.082 & 16.827 & 13.880 & 0.100 \\ 
 $^{16}$O &1.833 & 1.833 & 1.544 & 14.906 & 10.931 & 0.178 \\ 
 $^{18}$O &1.881 & 1.975 & 1.544 & 1.144  & 14.413 & 0.178 \\ 
 $^{23}$Na &2.773 & 2.81 & 0.54  & 3.546  & 4.887  & 0.336 \\ 
 $^{40}$Ca &3.51 & 3.43 & 0.563   &13.809 & 1.822  & 1.064 \\
 $^{44}$Ca &3.573 & 3.714 & 0.563 & 3.142 & 6.170  & 1.063 \\
 $^{75}$As &4.492 & 4.64 & 0.58   &0.353  & 1.688  & 2.624 \\
 $^{112}$Cd &5.38 & 5.58 & 0.58   & 2.075 & 4.462  & 4.861 \\
 $^{208}$Pb &6.624 & 6.890 & 0.549 &2.368 & 5.512  & 10.510\\ 
\hline \end{tabular}
\\[1mm] (*) The parameter $a$ is dimensionless for the
 MHO density form.
 \end{center} 
\caption{ \footnotesize Charge ($R_p, a$), neutron matter
 ($R_n, a$) density parameters,  $Q,\overline{Q}-$values and negative muon
 binding energies  for
 different nuclei. For carbon and oxygen we use a modified harmonic
 oscillator (MHO) density, $\rho(r) = \rho_0 (1+a(r/R)^2)\exp(-(r/R)^2)$,
 while for the rest of the nuclei, a two-parameter Fermi
 distribution, $\rho(r) = \rho_0  /(1+\exp((r-R)/a))$, was used. }  
\label{tab:dens} 
\end{table} 
\subsection{Inclusive Neutrino Reactions at Low Energies}

  In this subsection we present
results obtained by using non-relativistic kinematics for the
nucleons. We do not include FSI effects, since in
Sect.~\ref{sec:fsi} we made the approximation of setting to zero
Im$\Sigma$ for the hole states. For low nuclear excitation energies
($\le 60$), this approximation is not justified, because the imaginary
part of the selfenergy of particle and hole states are
comparable~\cite{FO92}. The inclusion of FSI effects would lead to a
quenching of the QE peak of the bare ph calculation, and a spreading
of the strength~\cite{GNO97,Bl01,Ma03,Am04,RPC}. However FSI effects
on integrated quantities are small. From the results of the
next subsection we estimate in $\sim 5$--10\% the theoretical error of
the integrated cross sections and total muon capture rates presented
in this subsection.

The processes studied in this subsection explore quite low nuclear
excitation energies ($\le$ 25--30 MeV), and hence one might expect
that a proper finite nuclei treatment could be in order. Indeed, these
processes are sensitive to the excitation of giant
resonances~\cite{Au02,Au97,Vol00,Ko03}. As mentioned in the
Introduction, our purpose is to describe the interaction of neutrinos
and antineutrinos with nuclei at higher energies (nuclear excitation
energies of the order of 100--600 MeV) of interest for future neutrino
oscillation experiments. However, our model provides a good
description of the low energy inclusive measurements analyzed in this
subsection. RPA correlations play an essential role and lead to
reductions as large as a factor of two.  We should remind that the
effective interaction appearing in the RPA series was fitted in
Ref.~\cite{Sp77} to giant resonances, and thus our approach
incorporates the mechanism which produces those resonances in finite
nuclei.

At low energies, finite nuclei effects are expected to be sizeable for
outgoing lepton energy distributions. There exist discrete and
resonance state peaks, and the continuum distribution significantly
differs from the LFG one.  However, the integrated strength over
energies, including the discrete state and resonance contributions,
remains practically unchanged, which explains the success of our model
to describe integrated-inclusive magnitudes. A clear example of this
can be found in Ref.~\cite{Am04} where the inclusive decay width of
muonic atoms by using a shell model with final neutron states lying
both in the continuum and in the discrete spectrum are calculated. The
results are compared with those obtained from a LFG model. Both
models\footnote{For simplicity in the calculations of
Ref.~\protect\cite{Am04}, RPA effects are not considered and the
static form of the nucleon CC current is employed.}  are in quite good
agreement within a few percent when the shell model density is used in
the LFG calculation. Being an integrated, inclusive observable, the
total capture width is quite independent of the fine details of the
nuclear wave functions. Similar conclusions were reached in the study
of the radiative pion capture in nuclei, $\left (A_Z-\pi^- \right
)_{\rm bound} \to \gamma + X $, performed in Ref.~\cite{RPC}.  There,
the predictions of a continuum shell model were also extensively
compared to those deduced from a LFG picture of the nucleus. The
differences found, among the integrated decay widths predicted by both
approaches, were, at most, of the order of 4\% (see Table 5 of first
entry in Ref.~\cite{RPC}).
\begin{figure}
\centerline{\includegraphics[height=20cm]{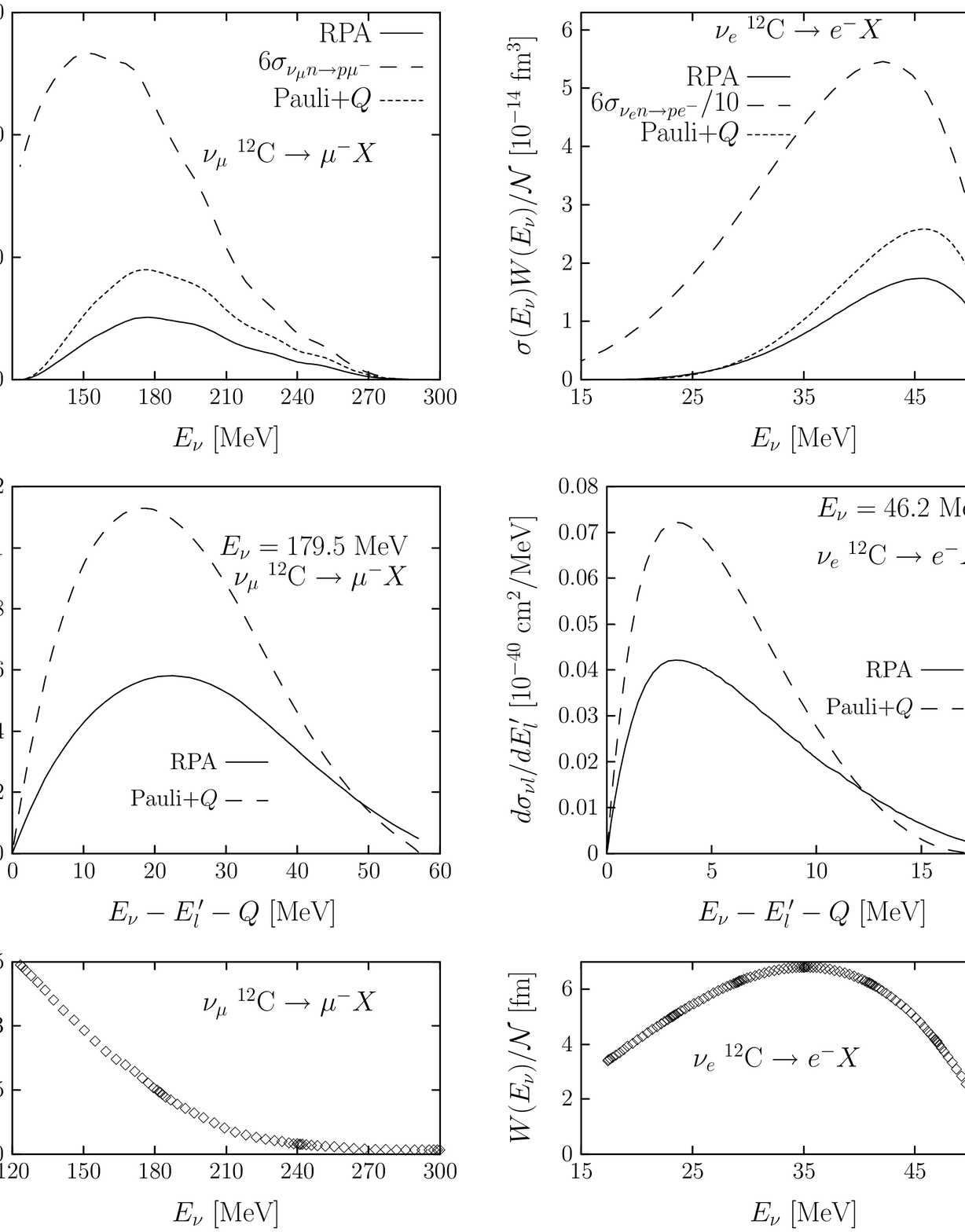}}
\caption{ \footnotesize Predictions for the LSND measurement of the
$^{12}$C $(\nu_\mu,\mu^-)X$ reaction (left panels) and the $^{12}$C
$(\nu_e,e^-)X$ reaction near threshold (right panels). Results have
been obtained by using non-relativistic kinematics for the nucleons
and without FSI. {\bf Top}: $\nu_\mu$ and $\nu_e$ cross sections multiplied
by the neutrino fluxes, as a function of the neutrino energy. In
addition to the RPA calculation (solid line), we show results without
RPA correlations and Coulomb corrections (dotted line), and also
(dashed line) the low density limit of Eq.~(\protect\ref{eq:ldt}).
{\bf Middle}: Differential muon and electron neutrino cross sections at
$E_{\nu_\mu}=179.5$ MeV (left) and $E_{\nu_e}=46.2$ MeV (right), as a
function of the energy transfer. {\bf Bottom}: Neutrino spectra 
from~\protect\cite{Al95} (left)  and Eq.~(\protect\ref{eq:fluxee}) (right).}
\label{fig:lsnd}
\end{figure}
\begin{table}
 \begin{center} \begin{tabular}{lccccccccc}\hline\tstrut 
   & LDT & Pauli+$Q$ & RPA & SM~\protect\cite{HT00} &
 SM~\protect\cite{Vol00} & CRPA~\protect\cite{Ko03}  
& \multicolumn{3}{c}{ Exp }
 \\\hline\tstrut &  &  &  &  &
  &  &
 LSND'95~\protect\cite{Al95}&LSND'97~\protect\cite{At97} &
 LSND'02~\protect\cite{Auexp02} \\ $\overline{\sigma}$
 $(\nu_\mu,\mu^-)$ & ~66.1~ & ~20.7~ & ~11.9~ & 13.2 &
 15.2 & 19.2 & ~$8.3 \pm 0.7 \pm 1.6$~ & ~$11.2 \pm 0.3 \pm 1.8$ ~&
 ~$10.6 \pm 0.3 \pm 1.8$~ \\ \hline\tstrut
 &  &  &  &  &  &  & 
 KARMEN~\protect\cite{KARMEN}&LSND~\protect\cite{LSND}  &
 LAMPF~\protect\cite{LAMPF} \\
$\overline{\sigma}$ $(\nu_e,e^-)$ 
& ~5.97~ & ~0.19~ & ~0.14~ &0.12 &
0.16 &0.15 & ~$0.15
 \pm 0.01 \pm 0.01$~ & ~$0.15 \pm 0.01 \pm 0.01$ ~& ~$0.141 \pm 0.023$~ \\
\hline
\end{tabular}
 \end{center} 
\caption{ \footnotesize Experimental and theoretical flux averaged
  $^{12}{\rm C}(\nu_\mu,\mu^-)X$ and $^{12}{\rm C}(\nu_e,e^-)X$ cross
  sections in 10$^{-40}$ cm$^2$ units. We label
  our  predictions  as in
  Fig.~\protect\ref{fig:lsnd}. We also quote results from other
  calculations (see text for details).} \label{tab:lsnd}
\end{table}
\subsubsection{The Inclusive Reactions $^{12}{\rm C}(\nu_\mu,\mu^-)X$
 and $^{12}{\rm C}(\nu_e,e^-)X$ Near Threshold}
In order to compare with the experimental measurements we calculate
flux averaged cross sections
\begin{equation}
\overline{\sigma}=
\frac{1}{{\cal N}}\int_{E_\nu^{\rm
min}}^{E_\nu^{\rm
max}}dE_\nu\sigma(E_\nu)W(E_\nu),\quad
 {\cal N}= \int_{E_\nu^{\rm
min}}^{E_\nu^{\rm
max}}W(E_\nu)dE_\nu,\label{eq:av}
\end{equation}
In the LSND experiment at Los Alamos, the inclusive
$^{12}{\rm}C(\nu_\mu,\mu^-)X$ cross section was measured using a pion
decay in flight $\nu_\mu$ beam, with energies ranging from zero to 300
MeV, and a large liquid scillantor
detector~\cite{Al95,At97,Auexp02}. The muon neutrino spectrum,
$W(E_\nu)$, is taken from Ref.~\cite{Al95} and it is plotted in the
left bottom panel of Fig.~\ref{fig:lsnd}.  We fix $E_\nu^{\rm min}$
and $E_\nu^{\rm max}$ to 123.1 and 300 MeV, respectively. The electron
neutrino beams used in experiments (LAMPF, KARMEN, etc.)  have relatively low
energies. Such neutrinos do not constitute a monochromatic beam, and
their spectrum\footnote{It is approximately described by the Michel
distribution
\begin{equation}
W(E_\nu) \propto E^2_\nu (E_\nu^{\rm max}-E_\nu),\quad
E_\nu^{\rm max} = \frac{m_\mu^2-m_e^2}{2m_\mu}, \quad E_\nu^{\rm
  min}=0
 \label{eq:fluxee}
\end{equation}}
is plotted in the right bottom panel of Fig.~\ref{fig:lsnd}. The bare
ph strength spreading due to the FSI might affect the inclusive, flux
averaged cross-section because of the energy variations in the
neutrino flux.  As an illustration, if some of the strength is shifted
to higher energies then some of the low-energy neutrinos will not be
able to excite it, compared with the case when the strength is not
spread out. Of course these effects are not very large, because some
strength is also moved to lower energies and compensates this. These
uncertainties contribute to the $\sim 5$--10\% theoretical error mentioned
above.

Our results for the $^{12}{\rm C}(\nu_\mu,\mu^-)X$ and $^{12}{\rm
 C}(\nu_e,e^-)X$ reactions near threshold are presented in
 Fig.~\ref{fig:lsnd} and Table~\ref{tab:lsnd}.  As can be seen in the
 table, the agreement to data is remarkable.  Nuclear effects turn out
 to be essential, and thus the simple prescription of multiplying by a
 factor of six (the number of neutrons of $^{12}$C) the free space
 $\nu_l n \to p l^-$ cross section overestimates the flux averaged
 cross sections by a factor of 5 and of 40 for the muon and electron
 neutrino induced reactions, respectively. The inclusion of Pauli
 blocking and the use of the correct energy balance in the reaction
 lead to much better results, but the cross sections are still badly
 overestimated. Only once RPA and Coulomb corrections are included a
 good description of data is achieved. RPA correlations reduce the
 flux averaged cross sections by about a factor of two, while Coulomb
 distortion significantly enhances them, in particular for the
 electron neutrino reaction where this enhancement is of about 30\%.

In Table~\ref{tab:lsnd}, a few selected
theoretical calculations (large basis shell model (SM) results of
Refs.~\cite{Vol00, HT00} and the continuum RPA (CRPA) ones from
Ref.~\cite{Ko03}) are also quoted. Our approach might look simplified
with respect to the ones just mentioned, but in fact it is also an RPA
approach built up from single particle states of an uncorrelated Fermi
sea. This method in practice is a very accurate tool when the
excitation energy is sufficiently large such that relatively many
states contribute to the process. Obviously, because of its nature,
the method only applies to inclusive processes and it is not meant to evaluate
transitions to discrete states. The adaptation of the method to finite
nuclei via the LDA has proved to be a rather precise technique to deal
with inclusive photonuclear reactions~\cite{CO92} and response
functions in electron scattering~\cite{GNO97}.  The
effective ph($\Delta$h)-ph($\Delta$h) interaction used in the RPA
series has been successfully employed in different
processes~\cite{GNO97,CO92,pion}. There are two distinctive features
of this interaction in the $S=T=1$ channel, which are not incorporated
in most of the finite nuclei approaches: i) it incorporates explicit
pion and rho exchanges and thus the force in this channel is splited
into longitudinal and transverse parts, and ii) it includes resonance
$\Delta$ degrees of freedom.  The inclusion of $\Delta$h components in
the RPA series reduces the LSND flux averaged $^{12}$C
$(\nu_\mu,\mu^-)X$ cross section by about a 15\%, while the reduction
factor is about four times smaller for the electron neutrino reaction,
because in this latter case, the larger contributions to the flux
averaged $^{12}$C$(\nu_e,e^-)X$ cross section comes from very low
($\le 20$ MeV) nuclear excitation energies (see Fig.~\ref{fig:lsnd}).
In addition, a correct tensorial treatment of the RPA hadronic tensor
is also important, and it explains the bulk of the existing
differences between our results and those obtained in Ref.~\cite{Si92}
(see Subsect.~\ref{sec:rpa} for details). As a matter of example, in
Ref.~\cite{Si92}   a value of $16.7 \pm 1.4 \times
10^{-40}~{\rm cm}^2$ is predicted for the LSND flux averaged
$^{12}$C$(\nu_\mu,\mu^-)X$ cross section. This value is about a 40\%
higher than our result, despite of using quite
similar ph($\Delta$h)-ph($\Delta$h) effective
interactions. Differences are significantly smaller for the electron
neutrino flux averaged cross section, since this reaction is sensitive
to quite lower energies.

In the middle panels of Fig.~\ref{fig:lsnd}, we plot the outgoing
lepton energy distribution for an incoming neutrino energy near the
maximum of $\sigma(E_\nu)W(E_\nu)$ (top panels). We see in
these plots the range of energies transferred ($E_\nu-E^\prime_l-Q$)
to the daughter nucleus: 25-30 MeV for the muon neutrino reaction and
less than 10 MeV for the electron neutrino process. Finite nuclei
distributions will present some discrete state and narrow resonance
peaks, but the integrated strength over energies would not be much
affected though, as we have already discussed.

\subsubsection{Total Nuclear Capture Rates for Negative Muons}

\begin{table}
 \begin{center} \begin{tabular}{ccc|cc}\hline\tstrut 
  & Pauli+$\overline{Q}$ $[10^4\,s^{-1}]$ & RPA $[10^4\, s^{-1}]$ &
Exp $[10^4\, s^{-1}]$ & $\left(\Gamma^{\rm Exp} -\Gamma^{\rm Th}
\right )/\Gamma^{\rm Exp} $ \\\hline \tstrut $^{12}$C & 5.42 & 3.21 &
$3.78\pm 0.03$ & \phantom{$-$}0.15 \\ $^{16}$O & 17.56 & 10.41 &
$10.24\pm 0.06$ & $-0.02$ \\ $^{18}$O & 11.94 & 7.77 & $8.80\pm 0.15$
& \phantom{$-$}0.12 \\ $^{23}$Na &58.38 & 35.03 & $37.73\pm 0.14$ &
\phantom{$-$}0.07 \\ $^{40}$Ca &465.5 &257.9 &$252.5\pm 0.6 $ &
$-0.02$ \\ $^{44}$Ca &318 &189 & $179 \pm 4 $ & $-0.06$ \\ $^{75}$As
&1148 & 679 & 609$\pm 4$ & $-0.11$ \\ $^{112}$Cd &1825 & 1078 &
1061$\pm 9 $ & $-0.02$ \\ $^{208}$Pb & 1939 & 1310 & 1311$\pm 8 $ &
\phantom{$-$}$0.00$ \\ \hline \end{tabular}
 \end{center} 
\caption{ \footnotesize Experimental and theoretical total muon
  capture widths for different nuclei. Data are taken from
  Ref.~\protect\cite{Su87}, and when more than one measurement is
  quoted in \protect\cite{Su87}, we use a weighted average:
  $\overline{\Gamma}/\sigma^2 = \sum_i \Gamma_i/\sigma_i^2$, with
  $1/\sigma^2 = \sum_i 1/\sigma_i^2$. Theoretical results have been
  obtained by using non-relativistic kinematics for the nucleons
  (Sect.~\ref{sec:nonrel} of the Appendix). To illustrate the role
  played by the RPA correlations, we quote two different theoretical
  results: i) Pauli+$\overline{Q}$ obtained from
  Eq.~(\protect\ref{eq:imc}) without including FSI effects and RPA
  correlations (i.e., replacing $A^{\mu\nu}_{\rm RPA}$ by $A^{\mu\nu}$
  in Eq.~(\protect\ref{eq:imc})), but taking into account the value of
  $\overline{Q}$; ii) the full calculation, including all nuclear effects
  with the exception of  FSI, presented in
  Sect.\protect\ref{sec:imc}, and denoted as RPA.  Finally, in the
  last column we show the relative discrepancies existing between the
  theoretical predictions given in the third column and data. }
\label{tab:capres} 
\end{table} 
\begin{figure}
\centerline{\includegraphics[height=15cm]{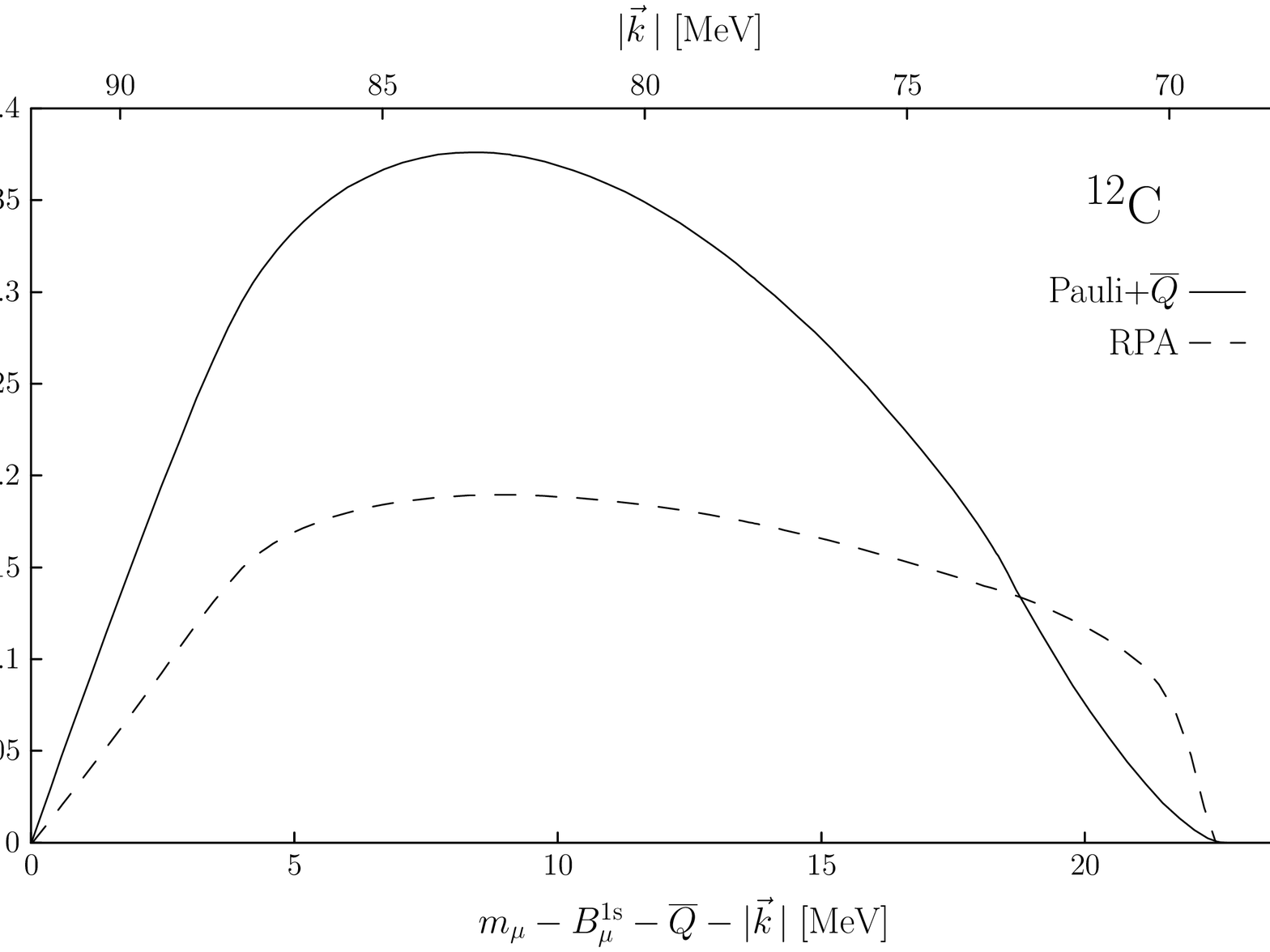}}
\vspace{-6cm}
\caption{ \footnotesize Inclusive muon capture differential width from
 $^{12}$C, as a function of the outgoing neutrino energy (top axis)
 and of the energy transfer (bottom axis). Non-relativistic kinematics
 has been used for the nucleons. The two  calculations are
 labeled as in Table~\protect\ref{tab:capres}.}\label{fig:cap}
\end{figure}
After the success in describing the LSND measurement of the reaction
$^{12}$C $(\nu_\mu,\mu^-)X$ near threshold, it seems natural to
further test our model by studying the closely related process of 
inclusive muon capture in $^{12}$C. Furthermore, and since there
are abundant and accurate measurements of nuclear inclusive muon
capture rates through the whole periodic table, we have also
calculated muon capture widths for a few selected nuclei, which will
be also studied below in Sect.~\ref{sec:altas}. Our results are
compiled in Table~\ref{tab:capres}. Data are quite accurate, with
precisions smaller than 1\%, quite far from the theoretical
uncertainties of any existing model.  Medium polarization effects
(RPA correlations), once more, are essential to describe the data, as was
already shown in Ref.~\cite{Ch90}.  Despite  the huge range of
variation of the capture widths\footnote{Note, $\Gamma^{\rm exp}$
varies from about 4$\times 10^4$ s$^{-1}$ in $^{12}$C to 1300 $\times
10^4$ s$^{-1}$ in $^{208}$Pb.}, the agreement to data is quite good
for all studied nuclei, with discrepancies of about 15\% at most. It
is precisely for $^{12}$C, where we find the greatest discrepancy with
experiment. Nevertheless, our model provides one of the best existing
combined description of the inclusive muon capture in $^{12}$C and the
LSND measurement of the reaction $^{12}$C $(\nu_\mu,\mu^-)X$ near
threshold~\cite{Ko03}. 

Finally, in Fig.~\ref{fig:cap} we show the outgoing  $\nu_\mu$ energy
distribution from muon capture in $^{12}$C, which ranges from 70 to 90
MeV. The energy transferred to the daughter nucleus ($^{12}$B)
ranges from 0 to 20 MeV. We also show in the figure the medium
polarization effect on the differential decay rate. As already
mentioned, the shape of the curves in Fig.~\ref{fig:cap} will
significantly change if a proper finite nuclei treatment is carried
out, with the appearance of narrow peaks, but providing  similar values
for the integrated widths~\cite{Am04}.

\subsection{Inclusive QE Neutrino and Antineutrino Reactions at
  Intermediate Energies}
\label{sec:altas}

\begin{figure}
\centerline{\includegraphics[height=23.5cm]{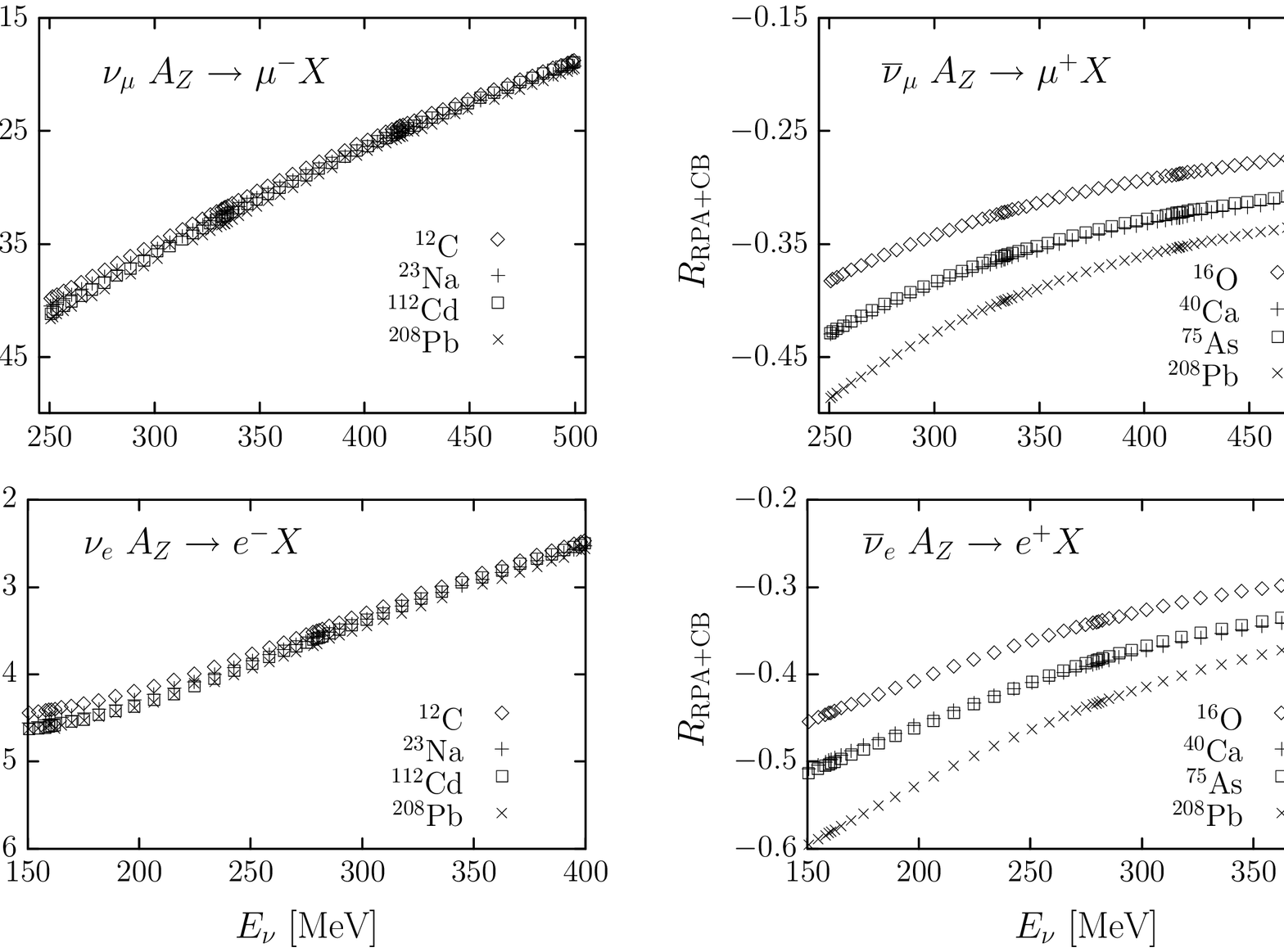}}
\vspace{-9cm}
\caption{ \footnotesize RPA and Coulomb (CB) corrections to electron
  and muon neutrino and antineutrino QE cross sections for different
  nuclei, as a function of the neutrino energy.  A  relativistic treatment of
  the nucleons is  undertaken and FSI effects are not considered. 
  $R_{\rm RPA+CB}$ is defined as 
  $(\sigma_{\rm RPA+CB}-\sigma_0)/\sigma_0$, where
  $\sigma_{0}$ does not include RPA  and Coulomb corrections,
  while $\sigma_{\rm RPA+CB}$  includes these 
   nuclear effects.  } \label{fig:corrections}
\end{figure}
\begin{figure}
\centerline{\includegraphics[height=23.5cm]{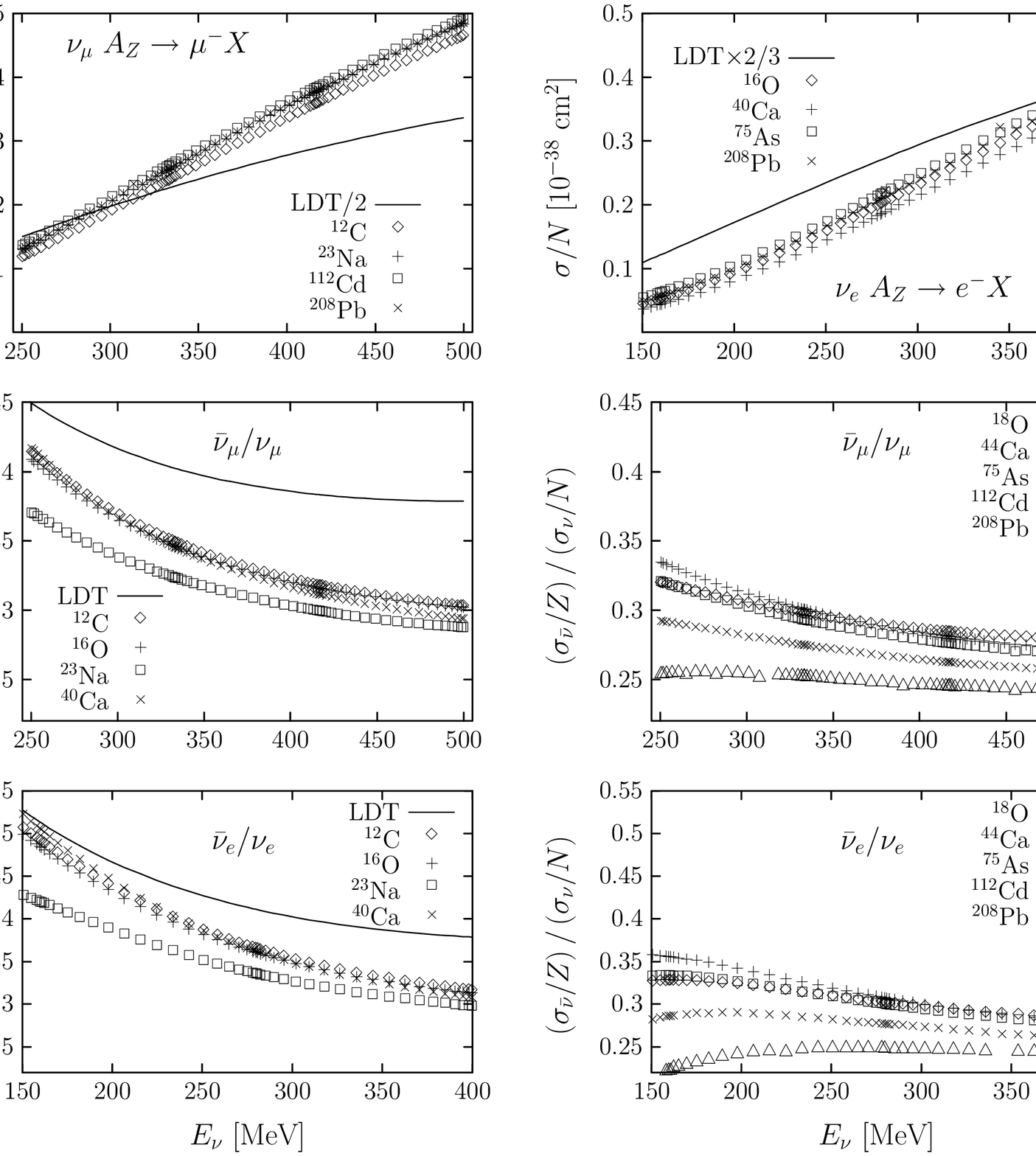}}
\vspace{-3cm}
\caption{\footnotesize Electron and muon neutrino and antineutrino
  inclusive QE cross sections and ratios for different nuclei, at
  intermediate energies. Results have been obtained with the full
  model without FSI, and using relativistic kinematics for the
  nucleons. For comparison we also show results 
  obtained in the free space (low density limit, LDT, 
  Eq.~(\protect\ref{eq:ldt})).}\label{fig:secs}
\end{figure}
In this subsection we will present results on muon and electron
neutrino and antineutrino induced reactions in several nuclei for
intermediate energies, where the predictions of the model developed in
this work are reliable, not only for integrated cross sections, as in
the previous subsection, but also for differential cross sections. We
will present results for incoming neutrino energies within the
interval 150-400 (250-500) MeV for electron (muon) species. The use of
relativistic kinematics for the nucleons leads to moderate reductions
of both neutrino and antineutrino cross sections, ranging these
reductions in the interval 4-9\%, at the intermediate energies
considered in this work. Such corrections do not depend significantly
on the considered nucleus.

In Fig~\ref{fig:corrections}, the effects of RPA and Coulomb
corrections are studied as a function of the incoming
neutrino/antineutrino energy. These corrections are important
(20-60\%), both for neutrino and antineutrino reactions, in the whole
range of considered energies. RPA correlations reduce the cross
sections, and we see large effects, specially at lower energies.  The
RPA reductions become smaller as the energy increases. Nevertheless for the
higher energies considered (500 and 400 MeV for muon and electron
neutrino reactions, respectively) we still find suppressions of about
20-30\%.  Coulomb distortion of the outgoing charged lepton enhances
(reduces) the cross sections for neutrino (antineutrino) processes.
Coulomb effects decrease with energy. For antineutrino reactions,
the combined effect of RPA and Coulomb corrections have a moderated 
dependence on $A$ and $Z$. Coulomb corrections reduce the
outgoing positive charged lepton effective momentum inside of the
nuclear medium. Thus, the phase space correction factor $|\vec{{\cal
K}}^{\,\prime}(r)| \hat{E}^\prime_l(r) /
|\vec{k}^{\,\prime}|E^\prime_l $ is smaller than one and the cross
section gets smaller. This effect, obviously grows with $Z$.  On the
other hand, the RPA suppression decreases when the lepton effective
momentum increases and it grows with $A$. The combined effect explains
the nuclear dependence found in the antineutrino plots. At the higher
energy end the $A-$dependence becomes milder, since Coulomb distortion
becomes less important. In the case of neutrinos, the increase of the
cross section due to Coulomb cancels out partially with the RPA
reduction. Finally, the existing differences between electron and muon
neutrino/antineutrino plots are due to the different momenta of an
electron and a muon with the same energy.

In Fig.~\ref{fig:secs}, we show electron and muon neutrino and
antineutrino inclusive QE cross sections and ratios for different
nuclei, as a function of the incoming lepton energy. Results have been
obtained with the full model presented in Sect.\protect\ref{sec:self},
including all nuclear effects with the exception of FSI, and using
relativistic kinematics for the nucleons. Neutrino cross sections
scale with $N$ (number of neutrons) reasonably well, while there exist
important departures from a $Z$ (number of protons) scaling rule for
antineutrino cross sections. These departures can be easily understood
from the discussion of Fig.\ref{fig:corrections}. To better
disentangle medium effects, the free space neutrino/antineutrino
nucleon cross section multiplied by the number of neutrons or protons
is also depicted in the plots.
\begin{figure}
\centerline{\includegraphics[height=23.5cm]{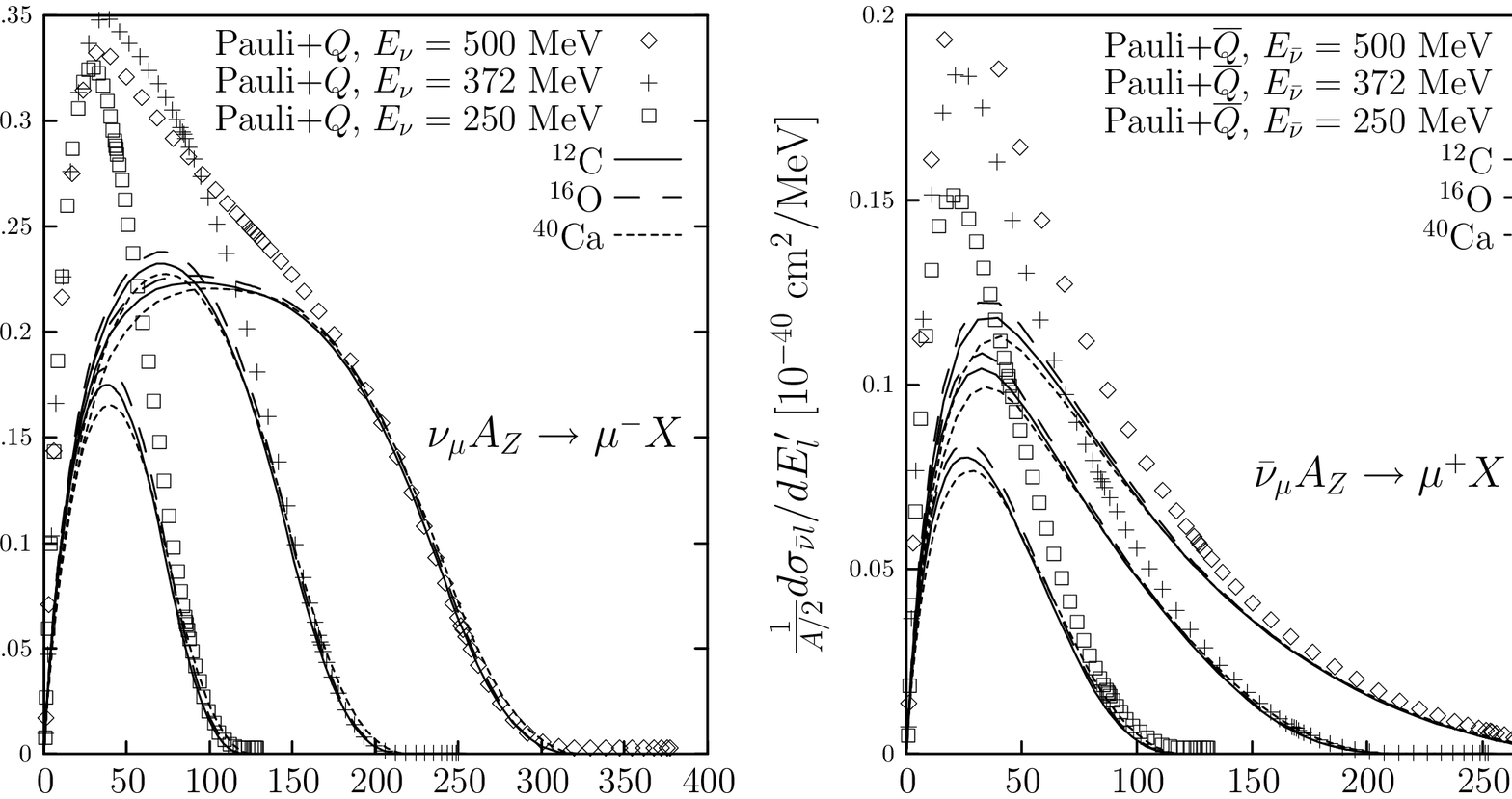}}
\vspace{-13cm}
\caption{\footnotesize Muon neutrino and antineutrino relativistic QE
  differential cross sections from different nuclei and several
  $\nu_\mu,{\bar \nu}_\mu $ energies. Results, denoted as 'Pauli+$Q$'
  or 'Pauli+$\overline{Q}$' have been obtained in $^{12}$C and do not
  include RPA, FSI and Coulomb effects, while the rest
  of results have been obtained with the full  model
  without FSI.}
  \label{fig:secdif}
\end{figure}

In Fig.~\ref{fig:secdif} we show muon neutrino and antineutrino
inclusive QE differential cross sections as a function of the energy
transfer, for different isoscalar nuclei and different incoming lepton
energies. We see an approximate $A-$scaling and once more the
important role played by the medium polarization effects. Similar
results (not shown in the figure) are obtained from electron species. 

\begin{figure}
\centerline{\includegraphics[height=17cm]{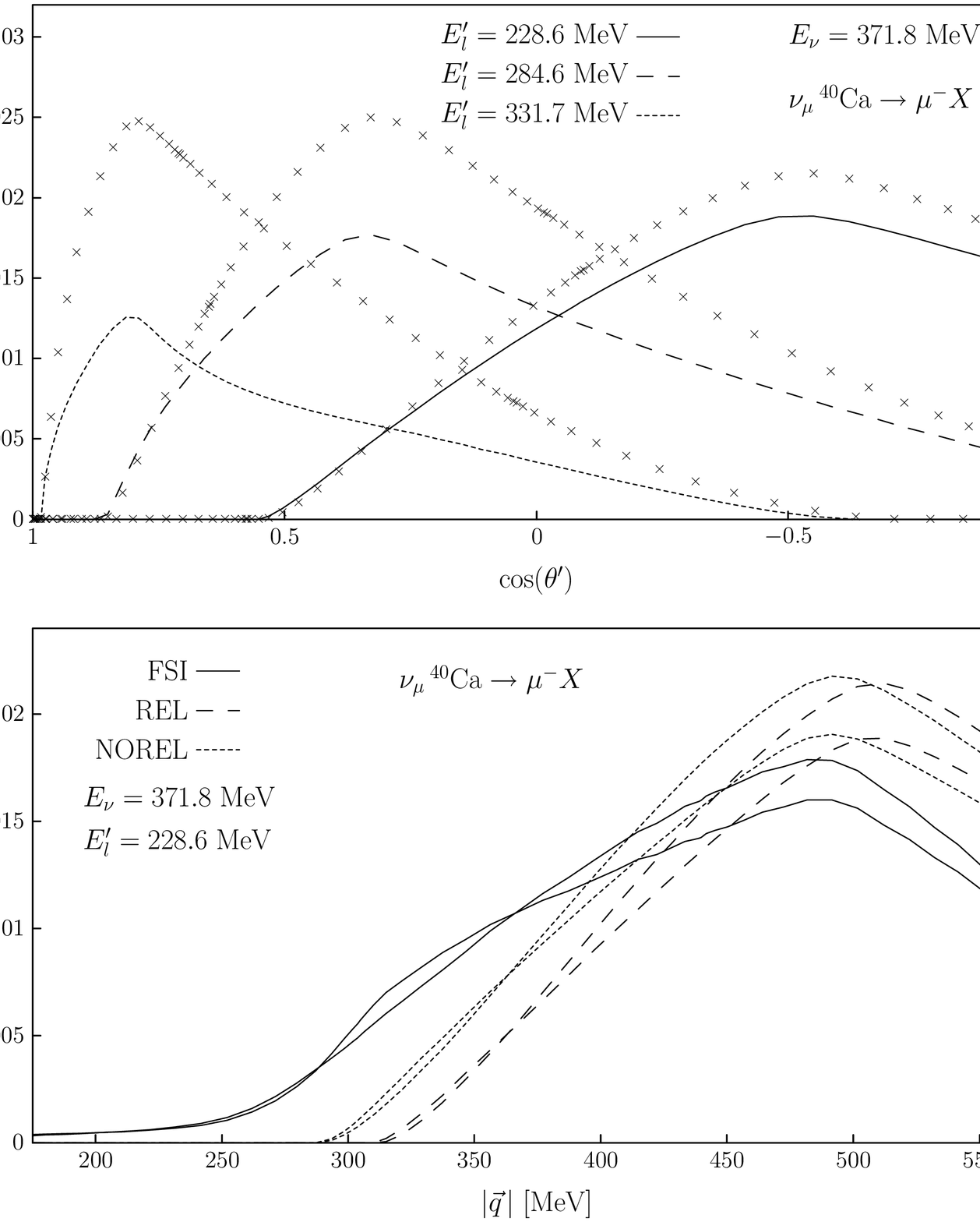}}
\vspace{-1cm}
\caption{\footnotesize Muon neutrino differential cross sections in
  calcium as a function of the lepton scattering angle (top) and of
  the momentum transfer (bottom). The neutrino energy is 371.8
  MeV. {\bf Top}: Cross sections, without FSI and using relativistic
  kinematics for the nucleons, at different muon energies. Crosses
  have been obtained without RPA and Coulomb effects, while the curves
  have been obtained with the full model (up to FSI effects).  {\bf
  Bottom}: Cross sections, for a muon energy of 228.6 MeV, obtained by
  using relativistic (long dashed line,'REL') and
  non-relativistic nucleon kinematics. In this latter case, we present
  results with (solid line , 'FSI') and without (short
  dashed line, 'NOREL') FSI effects. For the three
  cases, we also show the effect of taking into account RPA
  and Coulomb corrections (lower lines at the peak). The
  areas (in units of 10$^{-40}$ cm$^2$/MeV) below the curves are 3.50
  (REL), 3.87 (NOREL) and 3.77 (FSI) when RPA  and Coulomb
  corrections are not considered, and 3.13 (REL), 3.49 (NOREL) and
  3.53 (FSI) when these nuclear effects are taken into
  account.}\label{fig:secdouble}
\end{figure}

The double differential cross section
$d\sigma/dE^\prime_ld|\vec{q}\,|$ for the muon neutrino reaction in
calcium is shown in Fig.~\ref{fig:secdouble}.  In the top panel, we
compare the lepton scattering angle distribution for three different
values of the energy transfer. As usual in QE processes, the peaks of the
distributions are placed in the vicinity of $|\vec{q}\,|=
\sqrt{2Mq^0}$. In the bottom panel, we show FSI effects on 
 the differential cross section for one of the energies
 ($E_l^\prime=228.6$ MeV) studied in
the upper panel. We also show the effects  of using
relativistic kinematics for the nucleons. As anticipated, FSI provides
a broadening and a significant reduction of the strength of the QE
peak. Nevertheless the $|\vec{q}~|$ integrated cross section is only slightly
modified (a reduction of about 2.5\% when RPA corrections are not
considered and only about 1\% enhancement when they are included).

\begin{figure}
\centerline{\includegraphics[height=15cm]{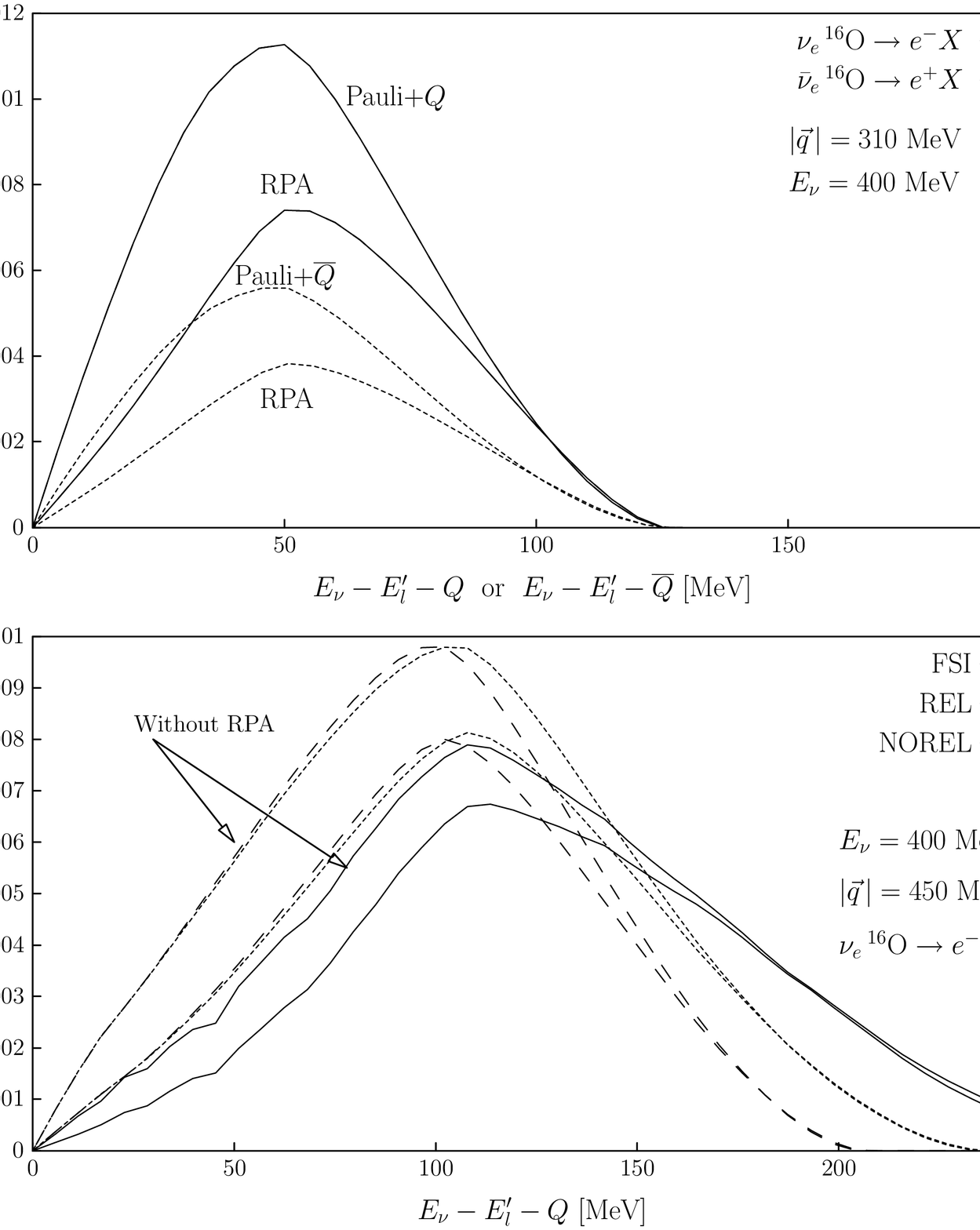}}
\vspace{-1cm}
\caption{\footnotesize $\nu_e$ and $\overline{\nu}_e$ differential
  cross sections in $^{16}$O as a function of the excitation energy,
  for fixed values of the momentum transfer and
  $E_{\nu,\overline{\nu}}=400$ MeV. {\bf Top:} Results obtained from
  the full relativistic model without FSI, with ('RPA') and without
  RPA and Coulomb corrections ('Pauli+$Q(\overline{Q})$').  {\bf
  Bottom:} Results obtained by using relativistic (long dashed line,
  'REL') and non-relativistic nucleon kinematics. In this latter case,
  we present results with (solid line, 'FSI') and without (short
  dashed line, 'NOREL') FSI effects. For the three cases, we also show
  the effect of taking into account RPA  and Coulomb
  corrections (lower lines at the peak). The areas (in units of
  10$^{-40}$ cm$^2$/MeV) below the curves are 1.02 (REL), 1.13 (NOREL)
  and 1.01 (FSI) when RPA and Coulomb corrections are not
  considered, and 0.79 (REL), 0.90 (NOREL) and 0.85 (FSI) when these
  nuclear effects are taken into account.  }
  \label{fig:dq}
\end{figure}

In Fig.~\ref{fig:dq} we plot double differential cross sections for
fixed momentum transfer, as a function of the excitation energy. We
show neutrino and antineutrino cross sections from $^{16}$O. FSI
effects are not considered in the top panel, and one finds the usual
QE shape, with peaks placed, up to relativistic corrections, in the
neighborhood of $\vec{q}^{\,2}/2M$. Once more, medium polarization
effects are clearly visible. FSI corrections are studied in the bottom
panel, and we find the expected broadening of the QE peak, but the
integrated cross sections remain almost unaltered.

\begin{figure}
\centerline{\includegraphics[height=17cm]{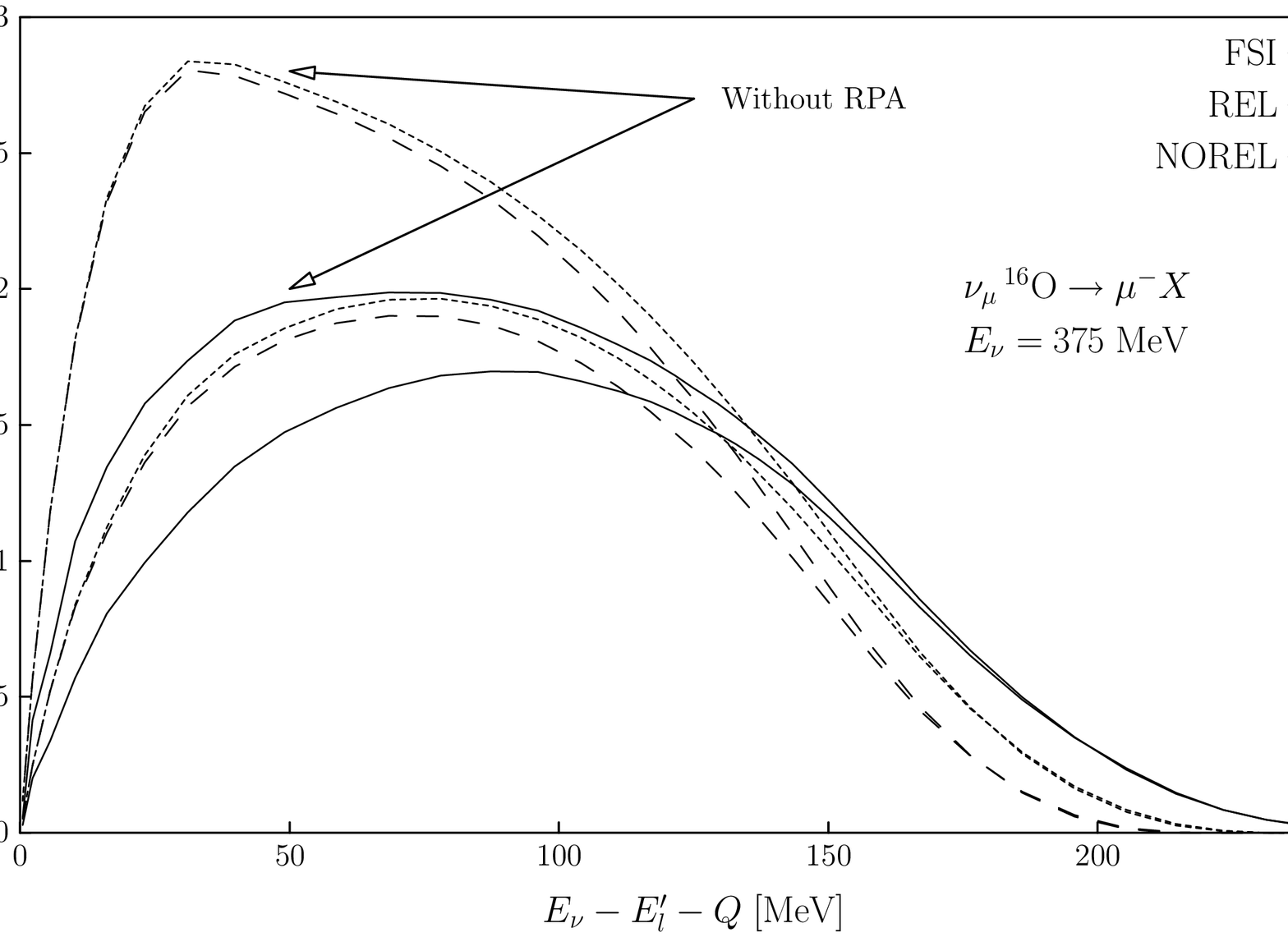}}
\vspace{-7cm}
\caption{ \footnotesize Muon neutrino  QE differential cross
  sections in oxygen as a function of the energy transfer.  The
  neutrino energy is 375 MeV. We show results for
  relativistic (long dashed line, 'REL') and
  non-relativistic nucleon kinematics. In this latter case, we present
  results with (solid line, 'FSI') and without (short
  dashed line, 'NOREL') FSI effects. We also show the effect of  RPA and Coulomb corrections (lower lines at the peak). The
  integrated cross sections can be found in  Table~\protect\ref{tab:fsi}. }\label{fig:fsi3}
\end{figure}

Finally, in Table~\ref{tab:fsi} we compile muon and electron neutrino
and antineutrino inclusive QE integrated cross sections from
oxygen. We present results for relativistic and non-relativistic
nucleon kinematics and in this latter case, we present results with
and without FSI effects. Though FSI change importantly the shape of
the differential cross sections, it plays a minor role when one
considers total cross sections. When medium polarization effects are
not considered, FSI provides significant reductions (13-29\%) of the
cross sections~\cite{Bl01}. However, when RPA corrections are included
the reductions becomes more moderate, always smaller than 7\%, and
even there exist some cases where FSI enhances the cross
sections. This can be easily understood by looking at
Fig.~\ref{fig:fsi3}, where we show the differential cross section as a
function of the energy transfer for $E_\nu=375$ MeV. There, we see
that FSI increases the cross section for high energy transfer. But for
nuclear excitation energies higher than those around the QE peak, the
RPA corrections are certainly less important than in the peak
region. Hence, the RPA suppression of the FSI distribution is
significantly smaller than the RPA reduction of the distribution
determined by the ordinary Lindhard function.
\begin{table}
 \begin{center} 
\begin{tabular}{cr|ccc||ccc}\hline\tstrut 
&&&&&&&\\
$E_\nu$ [MeV]&&\multicolumn{3}{c||}{$\sigma \left ( ^{16}{\rm O} (\nu_\mu,\mu^- X)
    \right )~[10^{-40} $
 cm$^2$]}&\multicolumn{3}{c}{$\sigma \left ( ^{16}{\rm O} ({\bar
      \nu}_\mu,\mu^+ X)\right )~[10^{-40} 
$ cm$^2$]} \\
&&&&&&&\\\hline\tstrut
 &          & ~~REL & ~~NOREL & FSI  &
  ~~REL & ~~NOREL & FSI \\\tstrut  
 500          & Pauli + $Q(\overline{Q})$~   &~~460.0&~~497.0  &431.6&~~155.8&~~168.4  &149.9\\\tstrut  
              & RPA                      ~   &~~375.5&~~413.0  &389.8&~~113.4&~~126.8  &129.7      \\\hline\tstrut  
 375          & Pauli + $Q(\overline{Q})$~   &~~334.6&~~354.8&292.2&~~115.1&~~122.6  &105.0 \\\tstrut  
              & RPA                      ~   &~~243.1&~~263.9  &243.9 &~~79.8 &~~87.9  &87.5    \\\hline\tstrut  
 250          & Pauli + $Q(\overline{Q})$~   &~~155.7&~~162.2  &122.5 &~~63.4 &~~66.4  &52.8    \\\tstrut  
              & RPA                      ~   &~~ 94.9&~~101.9  &93.6  &~~38.8 &~~42.1  &40.3    \\\hline\hline\tstrut  
&&&&&&&\\
$E_\nu$ [MeV]&&\multicolumn{3}{c||}{$\sigma \left ( ^{16}{\rm O} (\nu_e,e^-
    X)\right) ~[10^{-40} $
 cm$^2$]}&\multicolumn{3}{c}{$\sigma \left ( ^{16}{\rm O}({\bar
      \nu}_e,e^+ X )\right )~[10^{-40} 
$ cm$^2$]} \\
&&&&&&&\\\hline\tstrut
 &          &~~ REL & ~~NOREL & FSI  &
  ~~REL & ~~NOREL &  FSI \\\tstrut  
310           & Pauli + $Q(\overline{Q})$~   & ~~281.4    & ~~297.4 &240.6 &~~98.1 &~~ 104.0 & 87.2 \\\tstrut  
              & RPA                      ~   & ~~192.2    & ~~209.0 &195.2 &~~65.9 & ~~72.4  & 73.0   \\\hline\tstrut  
220           & Pauli + $Q(\overline{Q})$~   & ~~149.5    & ~~156.2 &121.2 &~~60.7 & ~~63.6  &51.0\\\tstrut  
              & RPA                      ~   & ~~ 90.1   & ~~ 97.3  &92.8  &~~36.8 & ~~40.0  &40.2    \\\hline\tstrut  
130           & Pauli + $Q(\overline{Q})$~   & ~~37.0    & ~~ 38.3  &28.8  &~~21.1 &~~21.9   &16.9\\\tstrut  
              & RPA                      ~   & ~~20.6    &  ~~22.3  & 23.3 &~~10.9 & ~~11.9  &12.8    \\\hline\tstrut  
\end{tabular}
 \end{center} 
\caption{ \footnotesize Muon (top) and electron (bottom) neutrino (left) and
  antineutrino (right)  inclusive QE  integrated cross sections from
  oxygen. We present results for relativistic ('REL') and
  non-relativistic nucleon kinematics. In this latter case, we present
  results with ('FSI') and without ('NOREL') FSI effects. Results, denoted
  as 'RPA' and 'Pauli+$Q(\overline{Q})$'  have been obtained  with and
  without including RPA correlations and Coulomb corrections, respectively.  }
\label{tab:fsi}
\end{table} 

\section{Conclusions} \label{sec:concl}

The model presented in this paper,
which is a natural extension of previous works~\cite{GNO97,
CO92,OTW82, pion} on electron, photon and pion dynamics in nuclei,
should be able to describe inclusive QE neutrino and antineutrino
nuclear reactions at intermediate energies of interest for future
neutrino oscillation experiments. Even though the scarce
existing data involve very low nuclear excitation energies, for which
specific details of the nuclear structure might play a role, our model
provides one of the best existing combined description of the
inclusive muon capture in $^{12}$C and of the measurements of the
$^{12}$C $(\nu_\mu,\mu^-)X$ and $^{12}$C $(\nu_e,e^-)X$ reactions near
threshold.  Inclusive muon capture from other nuclei is also
successfully described by the model.

The inclusion of RPA effects, in particular the nuclear
renormalization of the axial current, turned out to be extremely
important to obtain an acceptable description of data. This had been
already pointed out in Refs.~\cite{Ch90,Au02,Si92,Ko03}, and it is a
distinctive feature of nuclear reactions at intermediate
energies~\cite{GNO97, CO92,OTW82, pion}. On the other hand FSI
effects, though produce significantly changes in the shape of
differential cross sections, lead to minor changes for integrated cross
sections, comparable to the theoretical uncertainties,  once RPA
corrections are also taken into account.

The natural extension of this work is the study of higher transferred
energies to the nucleus, also relevant for the analysis of future
experiments aiming at determining the neutrino oscillation parameters
with high precision. For those energies, the production of real pions
and the excitation of the $\Delta(1232)$ or higher resonances will be
contributions to the inclusive neutrino--nucleus cross section
comparable to the QE one, or even larger.

\appendix

\section{CC Nucleon Tensor} 

\subsection {Impulse Approximation}
\label{sec:amunu}
Performing the traces in  Eq.~(\ref{eq:traamunu}) and taking into
account that in Eq.~(\ref{eq:res}) both the particle and the hole
nucleons are on the mass shell ($p^2=(p+q)^2 = M^2, ~ 2p\cdot q+q^2=0$), one
finds 
\begin{eqnarray}
A^{\mu\nu}(p,q)
&=& 16 (F_1^V)^2 \left \{ (p+q)^\mu p^\nu + (p+q)^\nu p^\mu
+ \frac{q^2}{2} g^{\mu\nu}\right \} 
+ 2q^2 (\mu_V F_2^V)^2 \left \{ 4 g^{\mu\nu} - 4\frac{p^\mu
p^\nu}{M^2} - 2\frac{p^\mu q^\nu + q^\mu p^\nu}{M^2} \right.  \nonumber \\
&-& \left. q^\mu q^\nu
 ( \frac{4}{q^2}+ \frac{1}{M^2} ) \right \} -
16F_1^V\mu_VF_2^V (q^\mu q^\nu-q^2 g^{\mu\nu})
+ 4G_A^2 \left \{ 2 p^\mu p^\nu + q^\mu p^\nu + p^\mu q^\nu +
g^{\mu\nu}(\frac{q^2}{2}-2M^2) \right. \nonumber \\
&-&\left.\frac{2M^2 (2m_\pi^2-q^2)}{(m_\pi^2-q^2)^2} q^\mu q^\nu
\right \}  - 16{\rm i} G_A\left (\mu_VF_2^V+F_1^V\right)
\epsilon^{\mu\nu\alpha\beta}q_\alpha p_\beta \label{eq:nucl}
\end{eqnarray}
The above tensor admits a  decomposition of the type
\begin{equation}
A^{\mu\nu}(p,q) = a_1 g^{\mu\nu} + a_2 \left ( p^\mu p^\nu +
\frac{p^\mu q^\nu + p^\nu q^\mu}{2}\right ) + {\rm i} a_3
\epsilon^{\mu\nu\alpha\beta}p_\alpha q_\beta + a_4 q^\mu q^\nu
\end{equation}
and  from Eq.~(\ref{eq:nucl}) we have
\begin{eqnarray}
a_1 (q^2) &=& 8q^2 \left \{ (F_1^V + \mu_V F_2^V)^2 + G_A^2\left (\frac14 -
\frac{M^2}{q^2}\right )  \right \} \nonumber \\
a_2 (q^2) &=& 32 (F_1^V)^2- 8 (\mu_V F_2^V)^2 \frac{q^2}{M^2} + 8
G_A^2 \nonumber \\  
a_3(q^2) &=&  16 G_A (F_1^V+\mu_V F_2^V) \nonumber \\ 
a_4(q^2) &=& - \frac{8q^2}{M^2}(\mu_V F_2^V)^2 \left (\frac{M^2}{q^2}+
\frac14\right )
-\frac{8M^2G_A^2}{m_\pi^2-q^2}\left(\frac{q^2}{m_\pi^2-q^2}+ 2\right)
 -16F_1^V \mu_V F_2^V \label{eq:aes}
 \end{eqnarray}

\subsection {RPA Corrections}
\label{sec:rpaamunu}
Taking $\vec{q}$ in the $z$ direction and after performing the RPA sum
of Fig.~\ref{fig:fig3}, we find, neglecting\footnote{Note that $q^0/M$
is of the order $|\vec{q}\,|^2/M^2$ and as mentioned in
Sect.~\ref{sec:rpa}, we have considered $\mu_V F_2^V|\vec{q}\,|/M$ of
order ${\cal O}(0)$.} corrections of order ${\cal
O}\left(k_F\vec{p}^{\,2}/M^2,k_F\vec{p}^{\, \prime 2}/M^2,k_Fq^0/M\right)$
\begin{eqnarray}
\frac{A^{00}_{\rm RPA}}{4M^2} &=& 8(F_1^V)^2 
\left \{\underline{\bf C_N}\,\left (\frac{E(\vec{p}\,)}{M}\right )^2
+\frac{q^2/4+q^0E(\vec{p}\,)}{M^2}  \right \}  - 2\frac{q^2}{M^2}
(\mu_V F_2^V)^2 \left \{  \frac{\vec{p}^{\,2}+q^0E(\vec{p})+ (q^0)^2/4}{M^2}
+ \frac{ (q^0)^2}{q^2} \right \}
\nonumber \\ 
&-&4 \underline{\bf C_N} F_1^V\mu_V F_2^V \frac{\vec{q}^{\,2}}{M^2} + 
 2 G_A^2 \left \{ \frac{q^0E(\vec{p}) + q^2/4+\vec{p}^{\,2}}{M^2}
-\underline{\bf C_L}\,\frac{(q^0)^2}{m_\pi^2-q^2}
\left(\frac{q^2}{m_\pi^2-q^2}+ 2\right) \right \} \\
&&\nonumber\\
\frac{A^{0z}_{\rm RPA}}{4M^2} &=&  4 (F_1^V)^2\left
\{\underline{\bf C_N}\frac{E(\vec{p}\,)}{M}\frac{2p_z+|\vec{q}\,|}{M}+
\frac{q^0p_z}{M^2}\right \} - \frac{q^2}{M^2}(\mu_V F_2^V)^2    \left \{
\frac{E(\vec{p}\,)}{M}\frac{2p_z+|\vec{q}\,|}{M}
 + 2\frac{q^0 |\vec{q}\,|}{q^2}
+ \frac{q^0 (2p_z+|\vec{q}\,|)}{2M^2} \right\} \nonumber\\
&-&4F_1^V \mu_V F_2^V \frac{q^0 |\vec{q}\,|}{M^2} 
+ 2G_A^2 \left \{
\underline{\bf C_L} \frac{E(\vec{p}\,)}{M}\frac{2p_z+|\vec{q}\,|}{2M}+ 
 \frac{q^0p_z}{2M^2} -\underline{\bf C_L}\, 
\frac{q^0 |\vec{q}\,|}{m_\pi^2-q^2}\left(\frac{q^2}{m_\pi^2-q^2}+
2\right)\right\}\\
&&\nonumber\\
\frac{A^{zz}_{\rm RPA}}{4M^2} &=& 8 (F_1^V)^2\left \{ 
 x\frac{p_z^2+|\vec{q}\,|p_z-q^2/4}{M^2} \right \}
-2\frac{q^2}{M^2}(\mu_V F_2^V)^2 \left \{ \left
(\frac{2p_z+|\vec{q}\,|}{2M}\right)^2 + \frac{(q^0)^2}{q^2} \right \}
-4\left (\frac{q^0}{M}\right)^2 F_1^V \mu_V F_2^V \nonumber\\ &+&
2G_A^2 \left \{ \underline{\bf C_L} +
\frac{p_z^2+|\vec{q}\,|p_z-q^2/4}{M^2} -\underline{\bf
C_L}\,\frac{\vec{q}^{\,2}}{m_\pi^2-q^2} \left(\frac{q^2}{m_\pi^2-q^2}+
2\right) \right \} \\ &&\nonumber\\
\frac{A^{xx}_{\rm RPA}}{4M^2} &=& 8 (F_1^V)^2 \left
\{\frac{p_x^2-q^2/4}{M^2} \right \} -2\frac{q^2}{M^2}(\mu_V F_2^V)^2 \left
\{\underline{\bf C_T}+\frac{p_x^2}{M^2} \right \} 
-4\underline{\bf C_T}\frac{q^2}{M^2}
F_1^V\mu_V F_2^V \nonumber\\
&+& 2G_A^2 \left \{
\underline{\bf C_T}  +
\frac{p_x^2-q^2/4}{M^2}\right\} \\ 
&&\nonumber\\
\frac{A^{xy}_{\rm RPA}}{4M^2} &=& 4{\rm i} G_A  (F_1^V+\mu_V F_2^V) 
\left (\frac{q^0 p_z}{M^2} 
-  \underline{\bf C_T} \frac{|\vec{q}\,|E(\vec{p}\,)}{M^2}\right ) 
\end{eqnarray}
with the polarization coefficients defined as
\begin{equation}
{\bf C_N}(\rho) = \frac{1}{|1-c_0f^\prime(\rho)U_N(q,k_F)|^2}, \quad
{\bf C_T}(\rho) = \frac{1}{|1-U(q,k_F)V_t(q)|^2}, \quad
{\bf C_L}(\rho) = \frac{1}{|1-U(q,k_F)V_l(q)|^2} \label{eq:coeffs}
\end{equation}
In order to preserve a Lorentz structure of the type $q^\mu q^\nu$,
for the pseudoscalar-pseudoscalar and pseudoscalar-axial vector terms
of the CC nucleon tensor, we have kept the RPA correction to the term
$\frac{(q^0)^2}{m_\pi^2-q^2} \left(\frac{q^2}{m_\pi^2-q^2}+ 2\right)$
in $A^{00}_{\rm RPA}$, despite of behaving like $\left
(q^0/|\vec{q}~|\right)^2 \approx {\cal O}(\vec{q}^{\,2}/M^2) $.

\section{ Basic Integrals} \label{sec:rel}

In a non-symmetric nuclear medium, the relativistic Lindhard function
is defined as
\begin{eqnarray}
\overline{U}_R(q,k_F^n,k_F^p) &=& 2  \int 
\frac{d^3p}{(2\pi)^3}\frac{M}{E(\vec{p})}\frac{M}{E(\vec{p}+\vec{q})}
 \frac{\Theta(k_F^n-|\vec{p}~|)
\Theta(|\vec{p}+\vec{q}~|-k_F^p)}{q^0+E(\vec{p})-
   E(\vec{p}+\vec{q})+{\rm i}\,\epsilon}  + (q \to -q)
\end{eqnarray}
The two contributions above correspond to the direct and the crossed
ph excitation terms, respectively. For positive transferred
energy only the direct term has imaginary part, which is given by
\begin{eqnarray}
{\rm Im}\overline{U}_R(q,k_F^n,k_F^p) &=& \int  d^3p {\cal
  F}_R(q,\vec{p}, k_F^n,k_F^p) = -M^2\frac{\Theta(q^0) 
\Theta(-q^2) }{2\pi|\vec{q}\,|}
\Theta(E_F^n-E_F^p+q^0)\Theta(E_F^n-{\cal E}_R^p) (E_F^n-{\cal  E}_R^p)
\end{eqnarray}
with 
\begin{eqnarray}
{\cal F}_R(q,\vec{p}, k_F^n,k_F^p) &=&-\frac{M^2}{4\pi^2}
\frac{\Theta(q^0)\delta(q^0 + E(\vec{p})
  -E(\vec{p}+\vec{q}~))}{E(\vec{p})E(\vec{p}+\vec{q})}
\Theta(k_F^n-|\vec{p}~|) \Theta(|\vec{p}+\vec{q}~|-k_F^p) \\
{\cal E}_R^p &=& {\rm Max}\left \{ M,E_F^p-q^0,
\frac{-q^0+|\vec{q}~|\sqrt{1-4M^2/q^2}}{2} \right \}, \quad E_F^{n,p}=
\sqrt{M^2 +(k_F^{n,p})^2},
\end{eqnarray}
being ${\rm Max}(...)$ the maximum of the quantities included in the bracket.
To perform the $d^3p$ integrations in Eqs.~(\ref{eq:res})
and~(\ref{eq:imc})  is
important to bear in mind that, though the LFG breaks down full Lorentz
invariance, one still has rotational invariance, thus we find
\begin{eqnarray}
T^0_R(q,k_F^n,k_F^p) &=&  \int 
d^3p {\cal  F}_R(q,\vec{p}, k_F^n,k_F^p)  E(\vec{p}) = 
\frac12 \left ( E_F^n + {\cal E}_R^p \right ) \times
  {\rm Im}\overline{U}_R(q,k_F^n,k_F^p)\\
\vec{T}_R(q,k_F^n,k_F^p) &=&  \int 
d^3p {\cal  F}_R(q,\vec{p}, k_F^n,k_F^p) \vec{p} 
= \left ( \frac{q^2}{2|\vec{q}~|^2}
{\rm Im}\overline{U}_R(q,k_F^n,k_F^p) +
\frac{q^0}{|\vec{q}~|^2} T^0_R(q,k_F^n,k_F^p) \right ) \vec{q}\\
R^{00}_R(q,k_F^n,k_F^p) &=& \int 
d^3p {\cal  F}_R(q,\vec{p}, k_F^n,k_F^p)  E^2(\vec{p}) 
= \frac13 \left ( (E_F^n)^2  +({\cal E}_R^p)^2+ {\cal E}_R^p
E_F^n\right ) 
\times  {\rm Im}\overline{U}_R(q,k_F^n,k_F^p)\\
\vec{R}_R(q,k_F^n,k_F^p) &=&  \int 
d^3p {\cal  F}_R(q,\vec{p}, k_F^n,k_F^p)  E(\vec{p}) \vec{p} =
  \left ( \frac{q^2}{2|\vec{q}~|^2} T^0_R(q,k_F^n,k_F^p)+
\frac{q^0}{|\vec{q}~|^2}R^{00}_R(q,k_F^n,k_F^p)  \right ) \vec{q} \\
R^{ij}_R(q,k_F^n,k_F^p) &=& 
\int 
d^3p {\cal  F}_R(q,\vec{p}, k_F^n,k_F^p) p^ip^j 
=  \frac{a_R-b_R}{2}\, \delta^{ij} +  
\frac{3b_R-a_R}{2|\vec{q}~|^2}\, q^iq^j, \qquad i,j=1,2,3 
\end{eqnarray}
with
\begin{eqnarray}
a_R(q,k_F^n,k_F^p) &=& R^{00}_R(q,k_F^n,k_F^p)-M^2 {\rm
  Im}\overline{U}_R(q,k_F^n,k_F^p) \\
b_R(q,k_F^n,k_F^p) &=& \frac{1}{4|\vec{q}~|^2} \Big \{ 
q^4 {\rm  Im}\overline{U}_R(q,k_F^n,k_F^p) + 
4 (q^0)^2 R^{00}_R(q,k_F^n,k_F^p)+4q^2q^0 T^0_R(q,k_F^n,k_F^p)  \Big \}
\end{eqnarray}

\section{ Non-relativistic Reduction of the Results of Appendix B 
 } \label{sec:nonrel}

We take a non-relativistic reduction of the nucleon dispersion
relation
\begin{equation}
E(\vec{p}) \approx  M + \frac{\vec{p}^{\,2}}{2M} \equiv{\bar E} (\vec{p})
\end{equation}
which implies, for consistency, that in the definition of the imaginary
part of the Lindhard function and in all integrals given in the
Appendix~\ref{sec:rel} the factors $M/E(\vec{p})$
and $M/E(\vec{p}+\vec{q})$ should be approximated by one. Thus, we
have\footnote{We suppress the subindex $R$ to distinguish the new
  expressions from the former ones.}
\begin{eqnarray}
\overline{U}(q,k_F^n,k_F^p) &=& 2  \int 
\frac{d^3p}{(2\pi)^3}
 \frac{\Theta(k_F^n-|\vec{p}~|)
\Theta(|\vec{p}+\vec{q}~|-k_F^p)}{q^0+{\bar E}(\vec{p})-
   {\bar E}(\vec{p}+\vec{q})+{\rm i}\,\epsilon}  + (q \to -q) 
\label{eq:lin_norel}
\end{eqnarray}
which correspond to the direct and the crossed 
ph excitation terms, respectively\footnote{For symmetric nuclear
  matter $\rho_p=\rho_n=\rho$, the above expression coincides, up to a
factor two due to isospin, with the definition of $U_N$ given in
Eq. (2.9) of Ref.~\protect\cite{Ga88}.}. For positive values of $q^0$ we have
\begin{eqnarray}
{\rm Im}\overline{U}(q,k_F^n,k_F^p) &=& \int d^3p {\cal F}(q,\vec{p},
  k_F^n,k_F^p) = -M^2\frac{\Theta(q^0) \Theta(-q^2)}{2\pi|\vec{q}\,|} 
\Theta({\bar E}_F^n-{\bar
  E}_F^p+q^0) \Theta({\bar E}_F^n-{\cal E}^p)  ({\bar E}_F^n-{\cal
  E}^p) 
  \label{eq:imu}
\end{eqnarray} 
with
\begin{eqnarray}
{\cal F}(q,\vec{p},
  k_F^n,k_F^p)&=& -\frac{1}{4\pi^2} \Theta(q^0)\delta(q^0 + {\bar E}(\vec{p})
-{\bar E}(\vec{p}+\vec{q}~))
 \Theta(k_F^n-|\vec{p}~|)
\Theta(|\vec{p}+\vec{q}~|-k_F^p) \\
{\cal E}^p & =& {\rm Max}\left \{ {\bar E}_F^p-q^0,
M + \frac{1}{2M}\left(\frac{Mq^0}{|\vec{q}\,|}-\frac{|\vec{q}\,|}{2}\right)^2
\right \}, \quad {\bar E}_F^{n,p}= M +\frac{(k_F^{n,p})^2}{2M}
\end{eqnarray}
To perform the integrations implicit in Eqs.~(\ref{eq:res})
and~(\ref{eq:imc})  we need 
\begin{eqnarray}
T^0(q,k_F^n,k_F^p) &=&  \int 
d^3p  {\cal F}(q,\vec{p},
  k_F^n,k_F^p)  {\bar E}(\vec{p}) = \frac12 
 \left ( {\bar E}_F^n + {\cal E}^p \right ) \times
  {\rm Im}\overline{U}(q,k_F^n,k_F^p)\\
\vec{T}(q,k_F^n,k_F^p) &=& \int 
d^3p  {\cal F}(q,\vec{p},
  k_F^n,k_F^p)  \vec{p}
=  \left (- \frac12 + \frac{Mq^0}{|\vec{q}\,|^2} \right ){\rm
  Im}\overline{U}(q,k_F^n,k_F^p)  \vec{q}\\
R^{00}(q,k_F^n,k_F^p) &=& \int 
d^3p  {\cal F}(q,\vec{p},
  k_F^n,k_F^p)  {\bar E}^2(\vec{p})= 
\frac13 \left ( ({\bar E}_F^n)^2  +({\cal  E}^p)^2+ {\cal E}^p{\bar E}_F^n
\right ) \times
  {\rm Im}\overline{U}(q,k_F^n,k_F^p)\\
\vec{R}(q,k_F^n,k_F^p) &=& \int 
d^3p  {\cal F}(q,\vec{p},
  k_F^n,k_F^p)  {\bar E}(\vec{p})\vec{p}
=  \left (\frac{Mq^0}{|\vec{q}\,|^2}-\frac12\right) 
T^0(q,k_F^n,k_F^p)~ \vec{q} \\
R^{ij}(q,k_F^n,k_F^p) &=& \int 
d^3p  {\cal F}(q,\vec{p},
  k_F^n,k_F^p) p^ip^j =  \frac{a-b}{2}\, \delta^{ij} + 
\frac{3b-a}{2|\vec{q}~|^2}\, q^iq^j, \qquad i,j=1,2,3 
\end{eqnarray}
with
\begin{eqnarray}
a(q,k_F^n,k_F^p) &=& 2M\Big\{T^0(q,k_F^n,k_F^p)-M\, {\rm
  Im}\overline{U}(q,k_F^n,k_F^p)\Big \}\\
b(q,k_F^n,k_F^p) &=& \frac{1}{4|\vec{q}~|^2} \left(2Mq^0  - |\vec{q}~|^2 
\right )^2 {\rm  Im}\overline{U}(q,k_F^n,k_F^p)
\end{eqnarray}

\section{Free Nucleon Cross Section}
  \label{sec:free} 

The cross section for the process $\nu_l +\, n \to l^- +
p $  is given by
\begin{equation}
\sigma_{\nu l} = \frac{G^2 \cos^2 \theta_c}{8\pi(s-M^2)^2}
\int_{q^2_{\rm min}}^{q^2_{\rm max}} dq^2 L_{\mu\nu}  A^{\nu\mu}\Big|_{
    p=(M,\vec{0})} \label{eq:free}
\end{equation}
where the leptonic ($L$) and nucleon ($A$) tensors are defined in
Eqs.~(\ref{eq:lep}) and~(\ref{eq:traamunu},\ref{eq:nucl}), respectively, 
$q^2_{\rm min (max)} = m^2_l - 2E_\nu(E^\prime_l \pm
|\vec{k}^\prime\,|)$ with $E_\nu$ and $E^\prime_l\,,\vec{k}^\prime$ 
the incoming neutrino LAB energy and outgoing lepton LAB energy and
momentum, and finally $s=(2E_\nu+M)M$. The variable $q^2$ is related
to the outgoing lepton LAB polar angle ($\theta^\prime$) 
by $q^2= (k-k^\prime)^2 = m^2_l - 2E_\nu(E^\prime_l -
|\vec{k}^\prime\,|\cos\theta^\prime)$. The tensor
contraction in Eq.~(\ref{eq:free}) gives in the LAB frame:
\begin{equation}
L_{\mu\nu}  A^{\nu\mu}\Big|_{
    p=(M,\vec{0})} = (q^2-m^2_l)\left\{ a_1  +
    \frac{s}{2}a_2 - \frac{q^2}{2} a_3 - a_4\frac{m^2_l}{2} \right \} + (s-M^2) \left \{ 
\frac{s-M^2}{2} a_2
    -q^2 a_3 \right \} \label{eq:free2}
\end{equation}
with the nucleon structure functions, $a_i(q^2)$,  given in Eq.~(\ref{eq:aes}).

The cross section for the process ${\bar \nu}_l +\, p \to l^+ +
n $  is obtained from  Eqs.~(\ref{eq:free}) and~(\ref{eq:free2}) by
replacing $a_3$ by $-a_3$.

\begin{acknowledgments}

J.N. warmly thanks to E.  Oset for various stimulating discussions and
communications.   This work  was  supported by  DGI  and FEDER  funds,
contract BFM2002-03218, and by the Junta de Andaluc\'\i a.

\end{acknowledgments}



\begin{thebibliography}{99}

\bibitem{GNO97} A. Gil, J. Nieves and E. Oset, Nucl. Phys. {\bf A627} (1997)
  543; {\it ibidem } Nucl. Phys. {\bf A627} (1997) 599.

\bibitem{Nuint04} See for instance, talks at ``The Third Workshop on
  Neutrino-Nucleus Interactions in the Few GeV Region {\it
  (NuInt04),http://nuint04.lngs.infn.it}~'', Gran Sasso, 
  2004.

\bibitem{Fu98} Y. Fukuda, et al., Phys. Rev. Lett. {\bf 81} (1998) 1562.

\bibitem{CO92} R.C. Carrasco and E. Oset, Nucl. Phys. {\bf A536} (1992) 445;
  R.C. Carrasco, E. Oset and L.L. Salcedo, Nucl. Phys. {\bf A541} (1992)
  585;  R.C. Carrasco, M.J. Vicente-Vacas and E. Oset,
  Nucl. Phys. {\bf A570} (1994) 701. 

\bibitem{OTW82} E. Oset, H. Toki and W. Weise, Phys. Rep.  {\bf 83} (1982) 281.

\bibitem{pion} L.L. Salcedo, E. Oset, M.J. Vicente-Vacas and
C. Garc\'\i a Recio, Nucl. Phys. {\bf A484} (1988) 557; C. Garc\'\i
a-Recio, et al., Nucl. Phys. {\bf A526} (1991) 685; J. Nieves,
E. Oset, C. Garc\'\i a-Recio, Nucl. Phys. {\bf A554} (1993) 509; {\it
ibidem} Nucl. Phys. {\bf A554} (1993) 554; E. Oset, et al.,
Prog. Theor. Phys. Suppl. {\bf 117} (1994) 461; C. Albertus,
J.E. Amaro and J. Nieves, Phys. Rev. Lett. {\bf 89} (2002) 032501;
{\it ibidem} Phys. Rev. {\bf C67} (2003) 034604.

\bibitem{nuint04} J. Nieves, J.E. Amaro and M. Valverde, {\it
  nucl-th/0408008}, talk given at
  ``The Third Workshop on Neutrino-Nucleus Interactions in the Few GeV
  Region'', Gran Sasso, 2004.

\bibitem{Pr59} H. Primakoff, Rev. Mod. Phys. {\bf 31} (1959) 802.

\bibitem{Mu77} N.C. Mukhopadhyay, Phys. Rep. {\bf 30} (1977) 1.


\bibitem{Gi81} N. Van Giai, N. Auerbach and A.Z. Mekjian,
  Phys. Rev. Lett. {\bf 46} (1981) 1444.


\bibitem{Na82} J. Navarro, J. Bernab\'eu, J.M.G. G\'omez and
  J. Martorell, Nucl. Phys. {\bf A375} (1982) 361.



\bibitem{Ch90} H.C. Chiang, E. Oset and P. Fern\'andez de C\'ordoba,
  Nucl. Phys. {\bf A510} (1990) 591; N.C. Mukhopadhyay, H.C. Chiang,
  S.K. Singh and E. Oset, Phys. Lett. {\bf B434} (1998) 7.
 
\bibitem{Ko94} E. Kolbe, K. Langanke and P. Vogel, Phys. Rev. {\bf C50}
  (1994) 2576; {\it ibidem} Phys. Rev. {\bf C62} (2000) 055502;

\bibitem{Ki85} C.W. Kim, S.L. Mintz, Phys. Rev. {\bf C31} (1985) 274.


\bibitem{HT00} A.C. Hayes and I.S. Towner, Phys. Rev. {\bf C61} (2000) 044603.


\bibitem{Au02} N. Auerbach, B.A. Brown,  Phys. Rev.{\bf C65} (2002) 024322.

\bibitem{Ja02} N. Jachowicz, K. Heyde, J. Ryckebusch and S. Rombouts,
  Phys. Rev. {\bf C65} (2002) 025501 

\bibitem{Ko03} E. Kolbe, K. Langanke, G. Mart\'\i
 nez-Pinedo and P.Vogel, J. Phys. {\bf G29} (2003) 2569.


\bibitem{Ga86} T.K. Gaisser and J.S. O'Connell, Phys. Rev. {\bf D34}
  (1986) 822.

\bibitem{Ku90} T. Kuramoto, M. Fukugita, Y. Kohyama and K. Kubodera,
  Nucl. Phys. {\bf A512} (1990) 711.

\bibitem{Si92} S.K. Singh and E. Oset, Nucl. Phys. {\bf A542} (1992)
  587;  {\it ibidem} Phys. Rev. {\bf C48} (1993)
  1246; T.S. Kosmas and E. Oset, Phys. Rev. {\bf 53} (1996) 1409;
  S.K. Singh, N.C. Mukhopadhyay  and E. Oset, Phys. Rev. {\bf C57}
  (1998) 2687.

\bibitem{Mi95} S.L. Mintz and M. Pourkaviani, Nucl. Phys. {\bf A594}
(1995) 346.

\bibitem{Um95} Y. Umino, and J.M. Udias,  Phys. Rev. {\bf C52} (1995) 
3399;  Y. Umino, J.M. Udias and  P.J. Mulders,  Phys. Rev. Lett. {\bf
  74} (1995) 4993. 

\bibitem{Ko97} E. Kolbe, K. Langanke,  F.K. Thielemann and P. Vogel,
 Phys. Rev. {\bf C52} (1995) 3437; E. Kolbe, K. Langanke and
 S. Krewald, Phys. Rev. {\bf C49} (1994) 1122; E. Kolbe, K. Langanke
 and P. Vogel, Nucl. Phys. {\bf A613} (1997) 382; {\it ibidem}
 Nucl. Phys. {\bf A652} (1999) 91.

\bibitem{Au97} N. Auerbach, N. Van Giai and O.K. Vorov,
  Phys. Rev. {\bf C56} (1997) R2368; N. Auerbach et al.,
  Nucl. Phys. {\bf A687} (2001) 289c.

\bibitem{Al97} W.M. Alberico et al., Nucl. Phys. {\bf A623} (1997)
 471; {\it ibidem} Phys. Lett {\bf B438} (1998) 9; {\it ibidem}
 Nucl. Phys. {\bf A651} (1999) 277.

\bibitem{Vol00} C. Volpe, et al., Phys. Rev. {\bf
  C62} (2000) 015501

\bibitem{Bl01} C. Bleve, et al., Astr. Part. Phys. {\bf 16} (2001) 145.

\bibitem{Ma03} C. Maieron, M.C. Martinez, J.A. Caballero and
J.M. Udias, Phys. Rev. {\bf C68} (2003) 048501.


\bibitem{MGP03} A. Meucci, C. Giusti and F.D. Pacati, Nucl.Phys. 
{\bf A739}, 277 (2004).

\bibitem{Gr04} K. M. Graczyk, {\it nucl-th/0401053}.

\bibitem{Ga71} S. Galster, et al., Nucl. Phys. {\bf B32} (1971) 221.

\bibitem{Sp77}  J. Speth, E. Werner and W. Wild,
 Phys. Rep. {\bf 33} (1977) 127; J. Speth, V. Klemt, J. Wambach and G.E. Brown 
 Nucl. Phys. {\bf A343} (1980) 382.

\bibitem{Wi74} D.H. Wilkinson, Nucl. Phys. {\bf A209} (1973) 470;
  Nucl. Phys. {\bf A225} (1974) 365.  

\bibitem{Ga88} C. Garc\'\i a-Recio, E. Oset and L.L. Salcedo,
  Phys. Rev. {\bf C37} (1988) 194.

\bibitem{En98} J. Engel, Phys. Rev. {\bf C57} (1998) 2004.

\bibitem{Be82} H. Behrens and W B\"uhring, {\it Electron Radial Wave
  Functions and Nuclear Beta Decay}, Clarendon, Oxford, 1982. 

\bibitem{Iz80} C. Itzykson and J..B Zuber, {\it Quantum Field Theory},
  McGraw-Hill, New York, 1980.

\bibitem{FO92} P. Fern\'{a}ndez de C\'{o}rdoba and E. Oset,
  Phys. Rev. {\bf C46} (1992) 1697.

\bibitem{Ra89} A. Ramos, A. Polls, and W, H. Dickhoff,
  Nucl. Phys. {\bf A503} (1989) 1

\bibitem{Mu95} H. M\"uther, G. Knehr and A. Polls, Phys. Rev. {\bf
  C52} (1995) 2955.

\bibitem{Sa88} L.L. Salcedo et al., Phys. Lett. {\bf B208} (1988) 339.

\bibitem{Ci90} C. Ciofi degli Atti, S. Liuti and S. Simula, Phys. Rev. 
{\bf C41} (1990) 2474.


\bibitem{Fi96} R.B. Firestone, {\it Table of Isotopes ($8^{\rm th}$
Edition)} , John Wiley \& Sons, 1996. 


\bibitem{Ja74} C.W. de Jager, H. de Vries and C. de Vries, At. Data
and Nucl. Data Tables {\bf 14} (1974) 479; 36 (1987) 495.

\bibitem{Ne75} J.W. Negele and D. Vautherin, Phys. Rev. {\bf C11}
(1975) 1031 and references therein.

\bibitem{GNO92} C. Garc\'\i a-Recio, J. Nieves and 
E. Oset, Nucl. Phys. {\bf A547} (1992) 473

\bibitem{Am04} J.E. Amaro, C. Maieron, J. Nieves and M. Valverde, {\it
  nucl-th/0409017}. 

\bibitem{RPC} J.E. Amaro, A.M. Lallena and J. Nieves, Nucl. Phys. {\bf
  A623} (1997) 529; H.C. Chiang et al., Nucl. Phys. {\bf A510} (1990)
  573, {\it erratum} Nucl. Phys. {\bf A514} (1990) 749. 

\bibitem{Al95} M. Albert, et al., Phys. Rev. {\bf C51} (1995) R1065.

\bibitem{At97} C. Athanassopoulos, et al., Phys. Rev. {\bf C56} (1997)
  2806.

\bibitem{Auexp02} L.B. Auerbach, et al., Phys. Rev. {\bf C66} (2002) 015501.

\bibitem{KARMEN} B. Zeitnitz, Prog. Part. Nucl. Phys. {\bf 32} (1994)
  351; B.E. Bodmann et al., Phys. Lett. {\bf B332} (1994)
  251.

\bibitem{LSND} C. Athanassopoulos, et al., Phys. Rev. {\bf C55} (1997)
  2078.

\bibitem{LAMPF} D.A. Krakauer et al., Phys. Rev. {\bf C45} (1992) 2450.


\bibitem{Su87} T. Suzuki, D.F. Measday and J.P. Roalsvig,
  Phys. Rev. {\bf C35} (1987) 2212 and
  references therein.

\end{thebibliography}
\end{document}